# Importance of Catalyst Stability *vis-à-vis* Hydrogen Peroxide Formation Rates in PEM Fuel Cell Electrodes

Vijay A. Sethuraman[1, 2, 4], John W. Weidner[2, *], Andrew T. Haug[3, 5], Marianne Pemberton[3], and Lesia V. Protsailo[3]

[2]Center for Electrochemical Engineering
Department of Chemical Engineering
University of South Carolina, Columbia, South Carolina 29208, USA
[3]UTC Power, South Windsor, Connecticut 06074, USA

The role of catalyst stability on the adverse effects of hydrogen peroxide ($H_2O_2$) formation rates in a proton exchange membrane fuel cell (PEMFC) is investigated for Pt, Pt binary (PtX, X = Co, Ru, Rh, V, Ni) and ternary (PtCoX, X = Ir, Rh) catalysts. The selectivity of these catalysts towards $H_2O_2$ formation in the oxygen reduction reaction (ORR) was measured on a rotating ring disc electrode. These measured values were used in conjunction with local oxygen and proton concentrations to estimate local $H_2O_2$ formation rates in a PEMFC anode and cathode. The effect of $H_2O_2$ formation rates on the most active and durable of these catalysts (PtCo and PtIrCo) on Nafion membrane durability was studied using a single-sided membrane electrode assembly (MEA) with a built-in reference electrode. Fluoride ion concentration in the effluent water was used as an indicator of the membrane degradation rate. PtIrCo had the least fluorine emission rate (FER) followed by PtCo/KB and Pt/KB. Though PtCo and PtIrCo show higher selectivity for $H_2O_2$ formation than unalloyed Pt, they did not contribute to membrane degradation. This result is explained in terms of catalyst stability as measured in potential cycling tests in liquid electrolyte as well as in a functional PEM fuel cell.

[1] – ISE Member
[4] – Present address: Environmental Energy Technology Division, Lawrence Berkeley National Laboratory, Berkeley, California 94720, USA.
[5] – Present address: 3M Center, 3M Fuel Center Components, St. Paul, Minnesota 55144, USA.
[*] – Corresponding author address: Professor, Department of Chemical Engineering, 3C05 Swearingen Engineering Center, University of South Carolina, 315 Main Street, Columbia, South Carolina 29208, USA.
Phone: + 1 (803) 777-3207; Fax: +1 (803) 777-8265; E-mail: weidner@engr.sc.edu

## 1. Introduction

Electrochemical oxygen reduction reaction (ORR) on Pt in acid media occurs as a four-electron transfer reaction resulting in $H_2O$ ($E^0$ = 1.23 V vs. SHE) and as a two-electron transfer reaction resulting in $H_2O_2$ ($E^0$ = 0.695 V vs. SHE) [1]. It is currently uncertain as to whether these two reactions occur in a parallel or in a serial fashion. The ratio of the extent of two electron transfer reaction to that of the four electron transfer reaction, the $H_2O_2$ selectivity, is a function of local water activity, and is typically between 0-20% for Pt [2]. The $H_2O_2$ selectivity is thought to have important consequences for PEM fuel cells from two perspectives – parasitic fuel loss as well as on the perceived role of $H_2O_2$ on membrane degradation [3,4,5,6,7]. As a result, almost all fuel cell catalyst groups study and report $H_2O_2$ selectivity as part of the characterization of novel fuel cell cathode catalysts.

The objective of this study is to evaluate the role of catalyst stability as well as the importance of $H_2O_2$ formation on Pt and Pt alloy catalysts. Both parasitic fuel loss due to $H_2O_2$ formation as well as its impact on durability is studied. In this article, the peroxide selectivity on binary and ternary Pt alloys during ORR is measured and the corresponding peroxide formation rates in a fuel cell anode and cathode are estimated. The most durable and active among the catalysts studied was evaluated in a single-sided MEA for their role in membrane degradation. The catalysts with the greatest selectivity for peroxide had the least membrane degradation and vice versa. This counter-intuitive result is explained in terms of their stability as measured in both liquid electrolyte cells as well as in an operating fuel cell.

## 2. Experimental
### 2.1. Catalyst Synthesis

Binary and ternary Pt alloys supported on Ketjen Black high surface area carbon (EC300J Carbon black, Akzo Nobel Polymer Chemicals, Chicago, IL) with approximately 50% Pt loading were synthesized according to the wet chemistry procedure [8]. In short, 2 g Ketjen Black powder as a carbon support was dispersed in deionized (DI) water, 5.75 g chloroplatinic acid ($H_2PtCl_6 \cdot 6H_2O$, Alfa Aesar) as a platinum salt was added to the dispersion and 5ml 37% formaldehyde (HCOH, Aldrich Chemicals) as a reducing agent was introduced afterward. In order to control the platinum particle size [9], carbon monoxide (20% CO in $N_2$, Praxair) was flowed to the dispersion when Pt was deposited on carbon. After being filtered, washed with DI water and dried in a vacuum oven at 90 °C, the Pt/C catalyst precursor was re-dispersed in a known quantity of the precursor solution according to the element needed to be alloyed (for example, Cobalt nitrate solution ($Co(NO_3)_2 \cdot 6H_2O$ for Co), and the mixture was dried at 90 °C. The dried mixture of Pt/C and the precursor compound was then heat-treated at 900 °C for 1 hr under a continuous flow of Argon (99.9%, Praxair). This step was to reduce metal ion to the corresponding metal by carbo-thermal reduction [8] and to alloy the Pt and the metal. This procedure is repeated with a corresponding precursor compound for synthesis of the ternary alloy. Detailed description of the synthesis can be obtained from [8] and [9]. Characterization of all of these binary and ternary alloys is carried out from a stability standpoint in this study.

All the catalysts reported in the reminder of this study are supported on Ketjen Black carbon (BET = 800 $m^2$ $g^{-1}$), unless noted otherwise. Pt/KB catalysts were synthesized in house and compared to two commercially available catalysts (Pt/KB and Pt/Vulcan XC-72R). The in-house Pt/KB catalyst performed as well as the commercially available ones and therefore the synthesized binary and ternary alloys were compared to commercially available Pt (TEC10E50E 46.7% Pt on Ketjen Black carbon, Tanaka Kikinzoku Kogyo KK, Japan) catalyst.

*2.2. Rotating Ring Disc Electrode (RRDE)*

The binary and ternary Pt alloys were evaluated for their selectivity towards $H_2O_2$ formation using the rotating ring disc electrode technique. Catalyst coated glassy carbon electrodes were prepared as described by Schmidt et al. [10]. Aqueous suspensions of 1 mg catalyst ml$^{-1}$ were obtained by pulse-sonicating 20 mg of the catalyst with 20 ml triple-distilled, ultrapure water (Millipore Corporation) in an ice bath (70% duty cycle, 60W, 15 minutes). Sonication was done using a Braun-Sonic U Type 853973/1 sonicator. A glassy carbon disc served as the substrate for the supported catalyst and was polished to a mirror finish (0.05 μm deagglomerated alumina, Buehler®) prior to catalyst coating. An aliquot of calculated amount of catalyst suspension was pipetted onto the carbon substrate, which corresponded to a Pt loading of ~14.1 μg Pt cm$^{-2}$. After evaporation of water for 30 minutes in $N_2$ atmosphere (15 in-Hg, vacuum), 14 μl of diluted Nafion solution (5% aqueous solution, 1100 EW; Solution Technology Inc., Mendenhall, PA) was pipetted on the electrode surface and further evaporated for 30 minutes in $N_2$ atmosphere (15 in-Hg, vacuum). Nafion® was used to adhere the Pt/Vulcan particles onto the glassy carbon electrode (the ratio of $H_2O$/Nafion® solution used was ca. 100/1). Previous work by Paulus et al. indicate that this procedure yielded a Nafion® film thickness of ca. 0.1 μm and that the utilization of the Pt/Vulcan catalyst (based on H-adsorption charge) on the electrode with this film was ~100%. In addition, the Nafion film rejects anions and only transports protons to the catalyst surface. It thus plays a role in minimizing anion adsorption on to the catalyst.

The catalyst-Nafion® coated electrode was immersed in deaerated (UHP Nitrogen, Praxair) Perchloric acid ($HClO_4$, 70%, ULTREX II® Ultrapure Reagent Grade, J. T. Baker) of varying concentrations for further synchronized chrono-amperometric and potentiodynamic experiments. Though a variety of supporting electrolytes are reported in the literature, anion adsorption on Pt is minimal for only a few electrolytes [11] (e.g., Trifluoromethane sulfonic acid (TFMSA) and $HClO_4$). In addition, the ultrapure reagent grade $HClO_4$ used in this study is free of ionic impurities; especially since $Cl^-$ ions, even in trace amounts (i.e. 1 ppm), is shown to drastically change both the activity and the reaction pathway of ORR on Pt catalysts [11, 12, 13]. All RRDE experiments were performed at atmospheric pressure and all solutions were prepared from ultrapure water (Millipore Inc., 18.2 MΩcm).

The electrochemical measurements were conducted in a standard electrochemical cell (RDE Cell®, Pine Instrument Company, NC) immersed in a custom-made jacketed vessel, temperature of which was controlled by a refrigerated/heating circulator (Julabo Labortechnik GMBH). A ring-disk electrode setup with a bi-potentiostat (Bi-Stat®, Princeton Applied Research Inc., TN) in conjunction with rotation-control equipment (Pine Instrument Company, NC). EC-Lab® software (version 8.60, Bio-logic Science Instruments, France) was used to control the bi-potentiostat. The Pt ring electrode was held at 1.2 V vs. SHE where the oxidation of peroxide is diffusion limited. The catalyst coated glassy carbon disc electrode (5 mm diameter, 0.1966 cm$^2$ area, DT21 Series, Pine Instrument Company, NC) was scanned between 0 – 1.2 V vs. SHE to characterize $H_2O_2$ formation within the potential range relevant to fuel cell operating conditions. Potentials were determined using a mercury-mercurous sulfate ($Hg/Hg_2SO_4$) reference electrode. All potentials in this study, however, refer to that of the standard hydrogen electrode (SHE). A high-surface area Pt cylindrical-mesh (5 mm diameter, 50 mm length) attached to a Pt wire (0.5 mm thick, 5 mm length) was used as the counter electrode.

*2.2.1. Effect of Oxygen Concentration*

The effect of oxygen concentration on ORR and $H_2O_2$ formation kinetics on Pt electrode was studied by varying the concentration of oxygen in the solution. The following three gases were used: oxygen (UHP grade, Praxair), Air (Industrial, Praxair) and 10.01% oxygen in nitrogen (Airgas). A gas flow meter (0-500 ml, Dwyer Instruments Inc., IN) was used to control the flow of the gas feed at ~100 ml min$^{-1}$ into the electrolyte. The electrochemical cell was sealed during the experiments to keep air from affecting the concentration of dissolved oxygen in the electrolyte. The concentration of dissolved oxygen in the electrolyte was estimated using the solubility values for oxygen in pure liquid water at 25 °C and 101 kPa [14].

*2.2.2. Effect of pH*

The effect of proton concentration on ORR and $H_2O_2$ formation kinetics on Pt, PtCo and PtIrCo catalysts was studied by varying the acidity of $HClO_4$ in the 2.0 – 0.1 M concentration window (~-0.301 – 1 pH, assuming $K_a \gg 1$ for $HClO_4$). Between solution changes, the electrochemical cell and its components were washed and boiled in DI water for 5 hours to ensure accurate pH levels. The catalyst-Nafion® coated electrode was also cleaned in a sonicator before every experiment with triple distilled ultrapure water.

*2.2.3. Collection Efficiency*

Standard procedure [15] for the determination of collection efficiency of a ring-disc electrode was followed. The electrodes were prepared as described above. The experiment was carried out in an electrochemical cell in deaerated (UHP Nitrogen, Praxair) 0.1 M $H_2SO_4$ (96.5%, J. T. Baker) with 10 mmol l$^{-1}$ $K_3Fe(CN)_6$ (99.7%, J. T. Baker. The disk electrode was swept at 1 mV s$^{-1}$ [vs. SHE] while the Pt ring was held at a constant potential of 1.2 V [vs. SHE]. At this ring potential, the oxidation of $[Fe(CN)_6]^{4-}$, produced at the disk electrode, to $[Fe(CN)_6]^{3-}$, proceeds under pure diffusion control. The collection efficiency was determined as N = $I_{ring}/I_{disk}$ = 0.20, which was independent of disk potential and consistent with the theoretical collection efficiency provided by the manufacturer of the ring-disc electrode [16].

*2.3. Potential Cycling – RDE Cell*

Potential cycling between 0.65 and 1.2 V vs. SHE was carried out in a standard RDE electrochemical cell at 25 °C and 1 atm in either 0.1 M H2SO4 or 0.1 M $HClO_4$. A high-surface area Pt cylindrical-mesh (5 mm diameter, 50 mm length) attached to a Pt wire (0.5 mm thick, 5 mm length) was used as the counter electrode. Platinized Pt electrode in contact with $H_2$ gas acted as the reference electrode. The electrode was held at each vertex potential for 5 seconds for a total of 20000 cycles. Electrochemical surface area measurements were conducted every 1000 cycles. A potentiostat (Princeton Applied Research Model 273A, Oak Ridge, TN) in conjunction with the Corrware software (Scribner Associates Inc., Southern Pines, NC ) was used for these measurements.

*2.4. Electrochemical Surface Area (ECA) Measurements*
*2.4.1. RDE Cell*

Cyclic voltammograms (CV) on the thin film RDE were recorded between 0 and 1.2V vs. SHE at 25 °C in deaerated 0.1 M $H_2SO_4$. The starting potential was the open circuit potential and scan rate was 5 mV s$^{-1}$. A typical CV consisted of three to five cycles, and showed very little cycle-to-cycle variation after the first cycle. For each catalyst studied, the CVs were recorded initially and after every 1000 cycles during the potential cycling test. The ECA was calculated from the charge under the voltammetric peaks corresponding to $H_2$ adsorption region in the CV after correcting for double layer charging (i.e., capacitive component) [17]. Mathematically, the ECA is given as,

$$\text{ECA} = \frac{Q_{Pt-H}}{\overline{Q}\,W} \qquad 1$$

In this equation, W corresponds to Pt loading in the catalyst layer in $mg_{Pt}$ cm$^{-2}$ (geometric area) and $\overline{Q}$, the charge density which is taken to be 210 μC cm$^{-2}$ (real area). This charge density corresponds to that of the Pt (100) facet [18, 19]. We assume this charge density to be valid for all the catalysts studied in this work. Finally, $Q_{Pt-H}$, the charge corresponding to $H_2$ adsorption, is calculated from,

$$Q_{Pt-H} = \frac{1}{v}\left(\int_{V_1}^{V_2} I dV - I_{dl}\Delta V\right) \qquad 2$$

The accuracy of this estimation depends on the identification of $V_1$ and $V_2$ in the CV, which are respectively, the potential where H adsorption begins and the potential where H coverage is complete before the rate of hydrogen evolution becomes significant. The term $I_{dl}\Delta V$ represents the double layer capacitance between these two potentials.

*2.4.2. Fuel Cell*

The CVs were recorded on the cathode after the cathode gas ($N_2$) was replaced with liquid water and the anode gas (pure $H_2$) was replaced with 4% $H_2$ in $N_2$. The CV parameters were similar to that of the RDE cell except the upper potential was 1 V. The ECA was estimated according to the above procedure.

*2.5. Single-sided Membrane Electrode Assembly with a Reference Electrode*

The effect of Pt alloy catalysts on membrane durability was studied using single sided MEAs designed with built-in reference electrodes. In order to specify the potential of the working electrode (anode or cathode), a reference electrode was placed inside the membrane in such a way that it was electronically isolated from the anode and the cathode but was in contact with $H_2$ gas. Since it is impossible to place a Pt or Au foil or wire inside a thin membrane, the reference electrode was sandwiched between two membranes. A perforated thinner membrane (Nafion® 111, in this case) electronically insulated the reference electrode from the anode Figure 1 shows the design of such an electrode. The working membrane in this case was thicker Nafion 117® membrane. For the reference electrode to work in a regular (two-sided) MEA, the flow-field should have an additional channel outside the projected electrode area so that the reference electrode has access to $H_2$. The same design as described above was used with electrode on only one side of the MEA. This design was an improvement over earlier durability experiments with single-sided MEA reported by Mittal et al. [20, 21, 22, 23]. In that, they built single sided MEAs (MEAs with catalysts and substrate on one side and no catalyst or substrate on the other side) they measured the fluorine emission rates at open circuit conditions  Their test matrix comprised of four such configurations with hydrogen and oxygen flowing on the anode and the cathode respectively and nitrogen flowing on either electrodes.  In this work, with the inclusion of the reference electrode, a potential can be applied upon the single sided electrode configuration.  The effect of potential on membrane durability can now be studied.  The effect of Pt, PtCo and PtIrCo catalysts on membrane degradation was studied at 0.6 V vs. SHE using these single-sided MEA setup with reference electrode.

*2.5.1. Fluorine Emission Rates (FER)*

Effluent water samples from the single-sided electrode experiment were collected in polyethylene bottles and analyzed for the presence and concentration of fluoride ions using a Dionex ICS-200 ion chromatography system.  In addition to this, a colorimetric method (Chemeterics Inc., VA) was used to measure the concentration of $H_2O_2$ in the effluent water.

*2.6. Fuel Cell Construction*

The PtCo, PtIrCo synthesized in house as well as commercially available Pt/C catalyst (TEC10E50E 46.7% Pt on Ketjen Black carbon [BET = 800 $m^2$ $g^{-1}$], Tanaka Kikinzoku Kogyo KK, Japan) were each mixed with Nafion® (5% aqueous solution, 1100 EW; Solution Technology Inc., Mendenhall, PA) ionomer and the resulting catalyst-ionomer ink (79% catalyst, 21% Nafion®) was coated onto Teflon® based (EI DuPont de Nemours & Company) decals.  These catalyst-coated decals were dried in $N_2$ at room temperature and atmospheric pressure for 30 minutes.  Catalyst-coated membranes (CCMs), 6.5 cm x 6.5 cm, were then made by hot pressing the catalyst-coated Teflon® decals onto both sides of Nafion® 112 membranes (proton form) at 130 °C and 4500 lbs for 300 seconds. Reinforced silicon, un-reinforced silicon and Teflon® pads were used as supports for this process.  While one side of the CCMs always had Pt catalyst, the other side had Pt, PtCo or PtIrCo catalyst.  The Pt, PtCo, and PtIrCo CCMs had similar Pt loadings of ~0.4 $mg_{Pt}$ $cm^{-2}$ on both the anode and the cathode sides.

The CCMs were each assembled into a fuel cell (25 $cm^2$ hardware, Fuel Cell Technologies Inc., NM).  Wet-proofed Toray® paper with micro-porous layer (243 μm total; Toray Industries, Japan) was used as gas diffusion media (GDM).  Gaskets were chosen in such a way that they allowed for 21% compression on the GDMs at 30 in-lbs of torque on the bolts.  The fuel cell had non-porous modular serpentine flow channels on the anode side and interdigitated flow channels (IDFF) on the cathode side.  The assembled fuel cell was tested for throughput, gas crossover and overboard leaks and then conditioned with $H_2$/$O_2$ (anode/cathode) in a fuel cell test stand (Habco Inc., CT) at 80 °C and 101 kPa (absolute).  Several current-voltage (VI) curves were measured in hydrogen and oxygen with 30 and 25% utilizations respectively until a steady high performance was reached.

*2.6. Potential Cycling – Fuel Cell*

Once the fuel cell reached a steady performance, the cathode gas was switched to $N_2$ and the temperatures on the humidity bottles were lowered such that the relative humidity was 50%. The cell temperature was increased to 120 °C and the cathode potential was cycled between 0.87 and 1.05V vs. SHE. The cathode remained at each potential for 1 minute each. The upper potential was lowered from 1.2 V (as was the case for the potential cycling in the RDE cell at 25 °C) to 1.05 V vs. SHE to avoid carbon corrosion, which could occur at 120 °C at higher potentials. . The potential cycling in the absence of $O_2$ captures the stability of the cathode catalyst whereas the potential hold experiment on the single-sided MEA captures the isolated effect of cathode peroxide formation on membrane durability. Electrochemical area (ECA) measurements were conducted at routine intervals. The ECA measurements on the cathode were done after replacing $N_2$ with liquid water. In addition to ECA measurements, performance curves were recorded with $H_2$/Air and $H_2$/$O_2$ at regular intervals. After 2800 cycles, the fuel cells were dismantled and XRD and EMPA analysis were done on the MEAs. X-ray diffraction data was used to analyze phase separation, crystal structure and particle size changes. Electron probe microanalysis was done to map the concentration of elements (i.e., Pt, Co or Ir) in the membrane.

*2.7. Fenton Tests*

Nafion® 111 membrane ($H^+$ form, DuPont Fluoroproducts, NC) samples measuring ~2" by ~2" were cut and heated in clean glass vials containing water at 80 ˚C for two hours to remove any surface impurities and solvents. The water was later drained from the vials and the membrane was dried in vacuum at 80 ˚C for four hours. The dry weight of the membrane was then measured. This weight was used to prepare various concentrations of aqueous ferrous sulfate ($FeSO_4.7H_2O$, 98.1% assay, Fischer Scientific Company, NJ) and cobalt nitrate ($Co[NO_3]_2.6H_2O$, reagent grade, Fischer Scientific Company, NJ) solutions such that the respective Fenton ion ($Fe^{2+}$ and $Co^{2+}$) uptake would vary. The dried Nafion® membrane was placed in these solutions for 15 hours in a $N_2$ atmosphere. This process impregnated various amounts of $Fe^{2+}$ and $Co^{2+}$ ions into the membrane samples. After this impregnation process, the membrane was dried in vacuum at 80 ˚C for four hours. The dry weight of the membrane impregnated with the Fenton ions was then measured.

The Fenton ion impregnated membrane samples were each placed in a Teflon® container containing 100 ml of 3% hydrogen peroxide aqueous solution (30%, VWR International) at 80 ˚C. Oxygen radicals are produced by the Fenton reaction, where the major step is $H_2O_2 + M^{2+} \rightarrow M^{3+} + HO^\bullet + HO^-$., where $M^{n+}$ represents a metal ion. The radical species attack the membrane via H abstraction, $HO^\bullet + RH \rightarrow H_2O + R^\bullet$ [3]. The radical R can further react to produce peroxyl and hydroperoxyl radical triggering a cascade of degradation reactions. Since hydrogen peroxide decomposes in the presence of a metal ion, the solution was replaced once every 24 hours and the leachate was saved for fluorine analysis. After 96 hours, the membrane was dried in vacuum at 80 ˚C for four hours. The weight loss, if any was recorded. The leachate was analyzed for fluorine using a Dionex ICS-200 ion chromatography system.

**3. Theory**

The ORR on Pt electrode occurs via the following reaction scheme,

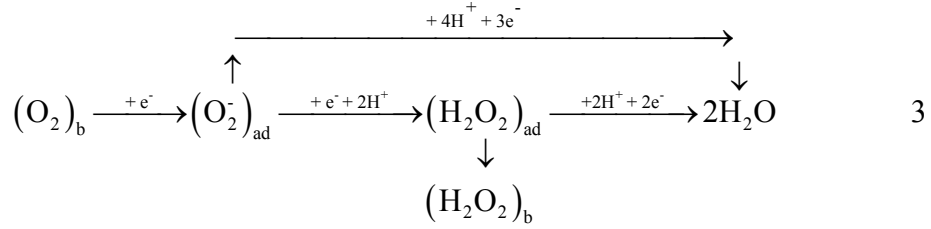

$$(O_2)_b \xrightarrow{+e^-} (O_2^-)_{ad} \xrightarrow{+e^-+2H^+} (H_2O_2)_{ad} \xrightarrow{+2H^++2e^-} 2H_2O \qquad 3$$

$$\downarrow$$

$$(H_2O_2)_b$$

It is currently uncertain whether the first electron transfer precedes the adsorption of $O_2$ or the adsorption of $O_2$ and the first electron transfer occur simultaneously. Nevertheless, the rate determining step appears to be the addition of the first electron to oxygen [24]. The second uncertainty is whether the splitting of the O-O bond occurs before the formation of the reaction intermediates (the direct 4e⁻ pathway) or whether the superoxo radicals react with the surrounding protons to form $H_2O_2$, which is further reduced to $H_2O$ (the serial 4e⁻ pathway) or escapes into the bulk.

The two-electron transfer reaction of $O_2$ reduction to $H_2O_2$, captured by the Pt ring, was analyzed in this study. At the ring, the $H_2O_2$ produced at the disk is oxidized back to $O_2$. The fraction of $H_2O_2$ formation, $\chi_{H_2O_2}$, can be determined from the collection efficiency, ring and disk currents by the expression,

$$\chi_{H_2O_2} = \frac{2I_{ring}/N}{I_{disk}+I_{ring}/N} \qquad 4$$

The measured current density j corresponding to $H_2O_2$ formation on a film covered RDE for the first-order ORR kinetics was previously reported to take the following expression [25], in terms of kinetic and mass-transport dependent currents,

$$\frac{1}{j} = \frac{1}{j_{kin}} + \frac{\delta_f}{nFD^f_{O_2}C^f_{O_2}} + \frac{1}{j_D} \qquad 5$$

Where j is

$$j = \frac{I_{ring}}{NA} \qquad 6$$

$j_{kin}$ is the current density in the absence of mass transfer effects and $j_D$ is the diffusion current given by the Levich equation.

$$j_D = 0.62nFD^{*\,2/3}_{O_2}C^*_{O_2}\nu^{-1/6}\omega^{1/2} \qquad 7$$

The concentration of $O_2$ in the solution was calculated from the partial pressure of $O_2$ in the inlet gas and $O_2$ solubility data for pure liquid water at corresponding temperature and 101 kPa [14]. The difference in $O_2$ solubility in pure liquid water and in $HClO_4$ (up to 2M) was assumed to be negligible. Combining equations 5 and 7 and solving for $j_{kin}$ gives,

$$j_{kin} = \frac{jnFD^f_{O_2}C^f_{O_2}D^{*\,2/3}_{O_2}C^*_{O_2}\omega^{1/2}}{nFD^f_{O_2}C^f_{O_2}D^{*\,2/3}_{O_2}C^*_{O_2}\omega^{1/2}-\delta_f jD^{*\,2/3}_{O_2}C^*_{O_2}\omega^{1/2}-1.6\nu^{1/6}jD^f_{O_2}C^f_{O_2}} \qquad 8$$

The purely kinetic portion of the $H_2O_2$ formation rate is

$$R_{H_2O_2} = \frac{j_{kin}}{2F} = k_f\left(C_{O_2}\right)^a\left(C_{H^+}\right)^b \qquad 9$$

Where,

$$k_f = k_f^0 \exp\left[\frac{\alpha F \eta}{RT^0}\right] \quad \quad 10$$

In equation 9, 'a' and 'b' are reaction orders with respect to $O_2$ and $H^+$ respectively. Only the forward rate term is used in equation 9 because at 0.6V vs. SHE and below, the rate of oxidation of $H_2O_2$ to $O_2$ is negligible. The kinetic rate constant $k_f$ was estimated for different potentials by plotting $H_2O_2$ production rate as a function of oxygen concentration for various potentials. Since the electrode reaction rate was earlier shown by Damjanovic and Hudson [26] to be faster on an oxide-free Pt surface than on an oxide-covered surface, both the forward and the reverse scans were used to estimate the reaction rate constant. The upper and lower bounds for the rate constants would therefore correspond to oxide-free and oxide-covered Pt surfaces respectively. The potential dependence of this rate constant is given in equation 10.

The activation energies for hydrogen peroxide formation reaction were evaluated by using the Arrhenius equation [27] shown below,

$$k_f^0 = k_{f,0}^0 \exp\left[\frac{E_a}{RT}\right] \quad \quad 11$$

This procedure is analogous to those described by Neyerlin et al. [28] and Bard and Faulkner [29]. The activation energies for $H_2O_2$ formation on supported Pt catalysts were compared to the computationally estimated activation energies reported in the literature. For example, using density functional theory (DFT), Anderson and Albu [30], Sidik and Anderson [31] and Wang and Balbuena [32] have reported activation energies for $H_2O_2$ formation on $Pt_1$, $Pt_2$ and $Pt_3$ sites respectively.

**4. Results and Discussion**
*4.1. $H_2O_2$ Selectivity, Activity and Stability of Binary and Ternary Pt Alloys*

The selectivity towards $H_2O_2$ formation for binary (PtX, X = Co, Rh, Ru, Ni and V) and ternary (PtCoX, X = Ir, Rh) alloy catalysts supported on high surface area Ketjen Black carbon are shown in Figure 2a and Figure 3a and respectively. The polarization curves for ORR on these binary and ternary alloys are shown in Figure 2b and Figure 3b and respectively. Data corresponding to the negative potential sweep (i.e., starting from 1.2 V vs. SHE and scanning towards 0 V vs. SHE) are shown in these figures. Figure 2a shows the selectivity of binary alloys compared to Pt. The selectivity values in the neighborhood of 0.6 V vs. SHE and at 0 V vs. SHE are of interest because of their relevance to fuel cell operating conditions. Normally, fuel cell cathode potentials are higher than the equilibrium potential for $H_2O_2$ formation (i.e., 0.695 V vs. SHE) except under very high load conditions when the local cathode potential could go negative of 0.695 V vs. SHE. On the anode, the local potential is close to the reversible hydrogen electrode potential. In general, the binary and ternary alloy catalysts show higher selectivity towards $H_2O_2$ formation than Pt. The PtRu catalyst is the only exception, which shows the least selectivity of these catalysts. As shown in Figure 2b, the disc current corresponding to PtRu does not show a characteristic point of inflexion near 0 V vs. SHE evident for all the other catalysts. The ring current (and thus the $H_2O_2$ selectivity), starts around 0.7 V vs. SHE for PtRu and reaches a maximum around 0.4V vs. SHE and then decreases monotonically until 0 V vs. SHE. Stamenkovic et al. [33] also observed similar results for PtRu and it is unclear as to why at potentials lesser than 0.1 V vs. SHE, PtRu is a better peroxide catalyst than pure Pt and most other binary and ternary Pt alloys. Clearly, in addition to Pt atoms Ru atoms are also involved in the cleaving of the O-O bond. For ORR on a pure Ru electrode in acid solution, see [34].

Among the binary alloys studied, PtNi shows the highest selectivity towards $H_2O_2$ formation followed by PtV and PtCo. Comparison of $H_2O_2$ selectivity of ternary alloys with Pt and PtCo is presented in Figure 3a. The stability of these binary and ternary alloys catalysts was evaluated by potential cycling experiments in an RDE cell. The results are summarized in Figure 4. The respective lattice constant for Pt is shown above each catalyst and a trend can be seen where the catalyst with lower lattice constant shows better stability The averaged crystallite size obtained via XRD for commercial Pt (TEC10E50E) and freshly synthesized PtRh, PtCo, PtIrCo and PtRhCo were 33, 31, 62, 59 and 35 Å, respectively. The true electrochemical area (ECA) corresponding to the charge under the voltammetric peaks for hydrogen desorption, corrected for double layer charging, is shown for every 1000 cycles for up to 20000 cycles. Qualitatively, all catalysts show ECA loss upon potential cycling. It can be seen that though Pt had the highest initial ECA , with ~85% of it lost after 20000 cycles. PtCo, PtRhCo and PtIrCo were the most stable among these catalysts losing 44.58%, 38.6%, and 44.94% of their initial ECAs respectively. The activity and durability of these catalysts relative to Pt is shown in Figure 5. Activity bars correspond to ratio of specific activity of a catalyst in terms of ORR current at 900 mV vs. SHE normalized to the ESA to that of Pt. Durability bars correspond to the ratio of ECA of a catalyst at the end of 20000 cycles to that of Pt. PtCo, PtRhCo and PtIrCo show the best activity and durability among the catalysts studied.

*4.2. $H_2O_2$ Kinetics and Formation Rates in a PEMFC Anode and Cathode*

The kinetics of electrochemical reduction of $O_2$ to $H_2O_2$ was studied in detail for PtCo and PtIrCo, the two catalysts that showed good initial specific activity and stability in the RDE tests. The rate constant, $k_f$, was measured from the kinetic portion of the ring current (Equation 9) for Pt, PtCo and PtIrCo and is shown in Figure 6. It shows the potential dependence of the rate constants according to Equation 10. The data between overpotential values 0-0.2V, 0.2-0.45V and 0.45-0.6V was fit with three separate linear equations. The resulting intercept, $k_f^0$, for these catalysts is summarized in Table 1. Since $k_f$ is a direct measure of the ring current, PtCo and PtIrCo show higher values than Pt.

The $H_2O_2$ formation rates measured as a function of water activity, potential and temperature using RRDE experiments was used to predict $H_2O_2$ formation rates at the anode and cathode of PEM fuel cell. Peroxide formation rate at the anode was predicted using oxygen permeability from the cathode and $\chi_{H_2O_2}$. Peroxide formation rate at the cathode was estimated via equation 9 , i.e. as a product of the rate constant and the local reactant concentrations. Peroxide formation at the cathode occurs only for fuel cells operating under considerable load (i.e., high cell current) such that the local potential goes negative relative to the equilibrium potential for peroxide formation. For estimation of cathode peroxide rates, local potential at the cathode was taken to be 0.6 V (i.e., η = 0.095 V).

Nafion® is a super-acid catalyst and hence the local acidity at the catalyst-membrane interfaces was calculated from the local water content and the fixed number of sulfonic acid groups.  The water sorption properties of Nafion® as a function of temperature and water activity had been studied by several laboratories.[35, 36, 37, 38, 39,40].  Using a novel tapered element oscillating microbalance (TEOM) technique, Jalani et al.[37] measured water uptake in Nafion as a function of water activity in vapor phase between 30 ºC and 110 ºC and reported that the water uptake increased with temperature and was highest at 110 ºC.  The difference in water uptake between 30 ºC and 110 ºC is negligible for lower water activities ($a_w < 0.7$).  This is reported by Jalani et al (experimental) and discussed in detail by Motupally et al. (simulations) [41].  For this work, the absorption isotherm of Nafion® 117 membranes measured at 30 °C by Zawodzinski et al.[38] were used.  Between water activity values of 0 and 1, the experimentally measured absorption isotherm was fit to the following polynomial [42],

$$\lambda = 0.043 + 17.81[a_w] - 39.85[a_w]^2 + 36.0[a_w]^3 \qquad 12$$

In this equation, $\lambda$ represents the number of water molecules per sulphonic acid group in the polymer and $a_w$ represents the activity of water, which is the effective mole fraction of water given by $p_0/p^*$, where $p^*$ is the vapor pressure of water, in bar. $p^*$ was calculated from the Antoine correlation,

$$\ln p^* = A_1 - \frac{B_1}{T+C_1} \qquad 13$$

The constants are $A_1 = 11.6832$, $B_1 = 3816.44$, $C_1 = -46.13$ [43].  Inside a fuel cell, this water activity is essentially the equilibrium relative humidity expressed as a fraction.  The concentrations of $H_2O$ and $H^+$ in the polymer are respectively expressed as,

$$C_{H_2O} = \frac{\rho \lambda}{EW} \qquad 14$$

$$C_{H^+} = \frac{C_{H_2O}}{\lambda} \qquad 15$$

In these equations, EW is the equivalent weight of the polymer (taken to be 1100) and $\rho$ is the humidity-dependent density of the polymer given by,

$$\rho = \frac{1.98 + 0.0324\lambda}{1 + 0.0648\lambda} \qquad 16$$

It was assumed that all sulphonic acid groups exist in a completely dissociated form.  Relating MEA acidity to water activity and hence to the humidity of the incoming gases facilitates in computing peroxide rates inside a fuel cell.  Quantitatively, the measured peroxide rates via the RRDE experiments at a particular oxygen concentration, pH value and temperature should equal the peroxide rates inside the fuel cell at same pH value and temperature.  Activation energies for $H_2O_2$ formation on supported Pt catalysts were estimated from kinetic currents obtained at 15 °C, 25 °C, 35 °C and 45 °C.  Oxygen permeability through Nafion® depends greatly on the water content of the membrane.  It has been shown by Sakai et al. [44] that $O_2$ diffusion rates in a completely dry Nafion® membrane has values similar to that in PTFE and approaches the limit of liquid water with increasing water content.  $O_2$ permeability was estimated using electrochemical monitoring technique (EMT) as a function of humidity and temperature and is comparable to those estimated by gas chromatography (GC) method [45].  Between 25% and 100% relative humidity of the feed gas, the permeabilities differ by as much as an order of magnitude.  Permeability for other temperatures and water contents were estimated by the following equation which was derived by fitting the measured permeability values,

$$P_{O_2}^m = \left(1.002 \times 10^{-14} - 9.985 \times 10^{-15} \, a_w\right) \exp\left[\left(0.0127 + 2.3467 \times 10^{-2} \, a_w\right) T\right] \quad 17$$

Oxygen solubility at the membrane-cathode catalyst layer interface, $C_{O_2}^c$, was estimated using the following relation,

$$C_{O_2}^c = \frac{P_{O_2}^m}{D_{O_2}^m} \quad 18$$

$D_{O_2}^m$ values for different temperatures and relative humidities were obtained from Sakai et al.'s work [44] and was fit to the following expression,

$$D_{O_2}^m = 9.78 \times 10^{-8} + 3.5 \times 10^{-9} T + 10^{-4} a_w \quad 19$$

Figure 7 shows the estimated peroxide rates at 75 °C for Pt, PtCo and PtIrCo catalysts at the cathode-membrane interface when the local cathode potential is 0.6 V and the gas feed is pure oxygen at 1 atm. The peroxide formation rates at the alloy cathodes are higher (~ three to four times) than that of Pt at all humidities. The estimated rates are similar to those experimentally measured by Liu and Zukerbrod [46]. The peroxide formation rate on Pt/Vulcan anode as a function of temperature and relative humidity is given in [2]. Estimated $H_2O_2$ formation rate on Pt/KB is lower than that on Pt/Vulcan. However, the estimated rates on both Pt/KB and Pt/Vulcan catalysts are lower than PtCo and PtIrCo catalysts supported on KB.

The potential profile across the membrane, measured in situ by Liu and Zuckerbrod [Figure 17 in Ref. 46] and modeled by Burlatsky et al.[47] at open circuit conditions, indicate that the potential at the anode-membrane interface is ~ 0 V. For the purpose of calculating $H_2O_2$ rates at the anode/membrane interface, a potential of ~0 V (i.e., η = 0.695 V) was assumed to exist at the interface.

The oxygen flux across the membrane from the cathode to the anode is,

$$F_{O_2} = \frac{D_{O_2}^m}{\delta}\left(C_{O_2}^c - C_{O_2}^a\right) \quad 20$$

The concentration of oxygen at the anode-membrane interface approaches zero, since all of the oxygen crossing over the membrane to the anode side is reduced to water or reacts chemically with hydrogen.

$$R_{H_2O_2}^a = \chi_{H_2O_2} \frac{P_{O_2}^m}{\delta} \quad 21$$

While the fraction of oxygen that is reduced to peroxide is a strong function of water activity, and is not a function of oxygen concentration [Figure 1c in Ref. 2]. An expression for $\chi_{H_2O_2}$ versus $C_{H^+}$ was obtained from measured values at room temperature for Pt, PtCo and PtIrCo,

$$\chi_{H_2O_2}^{Pt} = 0.2081 - 0.1208 \left(a_w\right) - 0.072 \left(a_w\right)^2 - 2.132 \times 10^{-14} \left(a_w\right)^3 \quad 22$$

$$\chi_{H_2O_2}^{PtCo} = 0.28 - 0.1625 \left(a_w\right) - 0.0967 \left(a_w\right)^2 \quad 23$$

$$\chi_{H_2O_2}^{PtIrCo} = 0.38 - 0.2206 \left(a_w\right) - 0.1313 \left(a_w\right)^2 - 2 \times 10^{-14} \left(a_w\right)^3 \quad 24$$

The estimated $H_2O_2$ formation rates at 75 °C for Pt, PtCo and PtIrCo at the anode-membrane interface as a function of relative humidity are shown in Figure 8. Again, the estimated peroxide formation rates are higher for PtCo and PtIrCo than Pt for all relative humidities. They go through a peak because oxygen permeability decreases with decreasing water activity whereas $H_2O_2$ selectivity increases with decrease in water activity. The estimated peroxide formation rates at 75 °C, 1 atm and 94% RH conditions for other binary and ternary catalysts are tabulated in Table 1. The parameter values used in this study are tabulated in Table 2. For a complete description of peroxide formation rates on Pt as a function of humidity and temperature, see [2].

The estimated $H_2O_2$ formation rate on the cathode is higher than that of the anode by about three orders of magnitude. This is in accordance with the observations of Panchenko et al. [48,49] who used electron paramagnetic resonance spectroscopy to detect the presence of peroxide generated radicals. According to them, the radical centers produced at the cathode side of the PEM fuel cell causes membrane degradation. Further, they did not observe any membrane degradation on the anode side. Though the presence of radical species in the vicinity of the cathode had been identified via spectroscopic methods, their formation rates as well as their concentration distribution and half-lives have not been quantified yet.

Parasitic fuel loss due to $H_2O_2$ formation was estimated as a ratio of molar rates of $H_2O_2$ formation in the anode and the cathode (both Pt) to that of $H_2$ and $O_2$ respectively, in a fuel cell operating at 2 A cm$^{-2}$, 75 °C and 94% RH fed with stoichiometric amounts of $H_2$ and $O_2$. The percent fuel loss is shown in Figure 9 for practical $H_2O_2$ selectivites. The loss of $H_2$ is less than a millionth of a percent while $O_2$ loss is less than a hundredth of a percent. Therefore, $H_2O_2$ selectivity as well as formation rates do not contribute to significant parasitic fuel loss.

*4.3. Role of PtCo and PtIrCo on Membrane Degradation*

Since the peroxide rates were higher for PtCo and PtIrCo than Pt and since these rates are higher at the cathode ($O_2$) than the anode ($H_2$), the effect of these catalysts on the membrane durability was tested on the oxygen side of a single-sided MEA with a built-in reference electrode. The reference electrode allowed specifying a potential on the cathode. In this circumstance, the reference electrode acted as the counter electrode as well. Figure 10 shows fluorine emission rates (FER) from the single-sided MEA experiments where the working electrode was held at 600 mV versus $H_2$ on Au reference electrode. The MEA with Pt showed the highest FER of the three electrodes. The initial FER for the Pt electrode (0.5 µmol hr$^{-1}$) is almost ten times higher than that of PtCo or PtIrCo electrodes. PtCo and PtIrCo electrodes have similar initial FER but PtCo shows more FER after 25 hours. Though Pt showed lower selectivity towards $H_2O_2$ formation than PtCo and PtIrCo catalysts, the MEA with Pt showed the most degradation. Similarly, MEA with the PtCo cathode showed higher degradation than that with the PtIrCo cathode. In effect, the catalyst with the highest peroxide selectivity and formation rates showed the least degradation rates and vice versa.

*4.4. Stability of PtCo and PtIrCo*

The above shown counter-intuitive result could be because of the extent of Pt and Co dissolution as $Pt^{2+}$ and $Co^{2+}$ and subsequent migration into the membrane, catalyzing membrane degradation reactions. To test this hypothesis, the stability of these catalysts was tested in a fuel cell via potential cycling between 0.87 and 1.05 V vs. SHE at 120 °C in the absence of $O_2$.

Figure 11 shows the elemental map of Pt, Co and Ir for the three MEAs with Pt, PtCo and PtIrCo cathodes after potential cycling between 0.87 and 1.05 V vs. SHE at 120 °C and 50% RH conditions. For elemental maps of MEAs before the potential cycling, see Figure 8 of Yu et al [50]. The cross-sections of the MEAs are shown with the cathode on the bottom. It is evident that the MEA with unalloyed Pt cathode shows the greatest concentration of Pt in the membrane among the three MEAs followed by those with PtCo and PtIrCo cathodes. Also, there is no evidence of Co or Ir in the membrane for the other two MEAs. The presence of elemental Pt at approximately one fifth of membrane thickness away from the cathode is due to the relative rates of diffusion between $H_2$ and $O_2$ (i.e., $D_{H_2}/D_{O_2} \cong 5$ ). The concentration of molecular $H_2$ and $O_2$ at this Pt plane tends to zero. The $Pt^{2+}$ ions dissolving from the cathode tends to reduce to Pt above this plane (i.e., on the anode side) in the presence of $H_2$ ($Pt^{2+} + H_2 \rightarrow Pt + 2H^+$).

These results indicate that Pt alloy catalysts minimize membrane poisoning that might occur during load cycling in a fuel cell. Co dissolution, albeit to lesser extent than Pt, is also evident. In that there exists a thin layer of free Co particles in the membrane along the cathode-membrane interface in the Co map for the PtCo cathode. XRD results on the post test MEAs (summarized in Table 3) showed significant sintering for the Pt cathode compared to PtCo and PtIrCo cathodes. For example, the Pt particle size increased from 3.3 nm to 6.1 nm for the Pt cathode. The particle size of PtIrCo changed from 5.9 nm to 6 nm with the final Pt particle size at 3.9 nm for the PtIrCo cathode. Phase separation was also observed during high temperature cycling. Post-test particle size analysis on the anode catalyst layer indicates sintering as well. Though all three anodes had the same catalyst (Pt/KB), the extent of sintering was different during the course of the potential cycling test and was dependent on the nature of the cathode catalyst. As shown in the last column in Table 3, anode Pt sintering decreases from Pt to PtCo to PtIrCo. This is the first time this has been observed and needs further exploring. The current-voltage curves recorded with $H_2/O_2$ as well as with $H_2$/Air at 120 °C and 50% RH (not shown) at routine intervals during the potential cycling showed no performance degradation for the cell with PtIrCo cathode compared to PtCo and Pt cathodes. Table 3b summarizes relative cyclic stability, Pt concentration in the membrane, particle size change and the extent of sintering for these three catalysts. For example, cyclic stability increases from Pt to PtCo to PtIrCo. For more on the performance and durability characteristics of PtIrCo catalysts, please see reports by Protsailo [51] and Haug et al. [52].

Figure 12 shows the normalized cathode ECA loss during potential cycling for these three catalysts. The trend seen here is qualitatively similar to that of the ECA loss seen after potential cycling experiment in the RDE cell. Pt cathode showed severe performance degradation with about 50% of its initial ECA lost after 2200 cycles. PtIrCo cathode showed very little degradation after similar number of cycles.

The apparent difference in the durability results (in terms of FER) in the single sided MEA experiment between PtCo and PtIrCo electrodes is explained by a Fenton test with $Co^{2+}$ as the Fenton ion. Figure 13 shows the observed weight loss from a Nafion® 111 membrane as a function of Fenton ion concentration for $Fe^{2+}$ and $Co^{2+}$ ions. It is readily perceivable that the presence of $Co^{2+}$ does degrade the membrane akin to $Fe^{2+}$, the well known Fenton's reagent. Fenton test with $Ir^{3+}$ as Fenton ion was not carried out because of the absence of iridium ions in the membrane even in the vicinity of the cathode catalyst-membrane interface as seen in the post cycling EPMA results. It should also be noted that results from a Fenton test alone does not dictate membrane durability under fuel cell conditions [53].

*4.5. Implications*

These results have implications on catalyst synthesis as well as on the operational aspects of a PEM fuel cell. $H_2O_2$ selectivity, on fuel cell catalysts (especially for cathodes) will now become less important compared to their stability. Consideration should be given if the $H_2O_2$ selectivity is abnormally high. For example, $H_2O_2$ selectivity of 100% will totally shut off the cathode reaction. Though the parasitic fuel loss for such high selectivity would be very small, the parasitic power loss would be 100%.

From a functional standpoint, this also impacts the practice of using air-bleed technique [54] (i.e., adding a small quantity of $O_2$ with the $H_2$ fuel) as a way to mitigate CO poisoning in the PEMFC anode. The current understanding is that air-bleed, in addition to CO oxidation, contributes to $H_2O_2$ formation in the anode and thus to membrane degradation [55]. This may not be true if the catalysts are more stable.

**5. Conclusions**

Binary and ternary Pt alloys supported on high surface area carbon were shown to have higher selectivity towards hydrogen peroxide formation in the ORR than unalloyed Pt. Though this higher selectivity meant higher peroxide formation rates in a functional fuel cell, the durability experiments indicated that MEAs with binary and ternary Pt catalysts had lower fluorine emission rates than those with just Pt, indicative that MEAs with alloys catalysts are more durable. This was because unalloyed Pt showed higher dissolution rates than binary and ternary Pt catalysts, which meant higher migration flux of $Pt^{2+}$ ions into the membrane. The presence of metal ions in the membrane in conjunction with the availability of gaseous $H_2$ and $O_2$ as well as peroxyl and hydroperoxyl radicals dictate membrane durability. Therefore, peroxide formation rates alone do not dictate the durability of an MEA. Further, based on estimated peroxide formation rates in an anode and a cathode of a functional fuel cell, the parasitic fuel loss was shown to be less than a millionth of a percent for $H_2$ and less than a thousandth of a percent for $O_2$ for practical $H_2O_2$ selectivites. Therefore, it is concluded that $H_2O_2$ selectivity and formation rates in fuel cell catalysts may not be as important as previously thought on both durability and parasitic fuel loss standpoints. We recommend based on these studies that mass activity and stability (i.e., low catalyst dissolution) be given more importance than $H_2O_2$ selectivity during the process of fuel cell catalyst design and characterization.

**List of Symbols**

| | |
|---|---|
| a | reaction-order with respect to $O_2$ in the $H_2O_2$ formation reaction |
| $a_w$ | water activity |
| A | disk area, $cm^2$ |
| b | reaction-order with respect to $H^+$ in the $H_2O_2$ formation reaction |
| $C_{H^+}$ | proton concentration, mol $cm^{-3}$ |
| $C_{H_2O}$ | water concentration in the membrane, mol $cm^{-3}$ |
| $C_{O_2}^*$ | oxygen concentration in the bulk of the electrolyte, mol $cm^{-3}$ |
| $C_{O_2}^f$ | oxygen concentration in Nafion® film, mol $cm^{-3}$ |
| $C_{O_2}^a$ | oxygen concentration in Nafion® 112 membrane-anode catalyst layer interface, mol $cm^{-3}$ |
| $C_{O_2}^c$ | oxygen concentration in Nafion® 112 membrane-cathode catalyst layer interface, mol $cm^{-3}$ |
| $D_{O_2}^*$ | oxygen diffusion coefficient in the electrolyte, $cm^2\ s^{-1}$ |
| $D_{O_2}^f$ | oxygen diffusion coefficient in Nafion® film, $cm^2\ s^{-1}$ |
| $D_{O_2}^m$ | diffusion coefficient of $O_2$ in Nafion® 112 membrane, $cm^2\ s^{-1}$ |
| $E^0$ | equilibrium potential, 0.695 V vs. SHE at 25 °C and 101 kPa |
| $E_a^*$ | activation energy for $H_2O_2$ formation, J $mol^{-1}$ |
| $E_{app}$ | applied potential, V vs. SHE |
| EW | equivalent weight of Nafion® polymer, 1100 g $equiv^{-1}$ |
| F | Faraday constant, 96485 C $mol^{-1}$ |
| I | current density, A $cm^{-2}$ |
| $I_{ring}$ | ring current, mA |
| $I_{disk}$ | disk current, mA |
| j | total peroxide current density, mA $cm^{-2}$ |
| $j_{disk}$ | disk current density, mA $cm^{-2}$ |
| $j_D$ | diffusion-limited current density, mA $cm^{-2}$ |
| $j_{kin}$ | kinetic current density, mA $cm^{-2}$ |
| $k_b$ | rate constant for $H_2O_2$ electro-oxidation, $s^{-1}$ |
| $k_f$ | rate constant for $H_2O_2$ formation, $mol^2\ cm^{-5}\ s^{-1}$ |
| N | collection efficiency |
| n | number of electrons transferred per $O_2$ molecule in $H_2O_2$ formation, 2 |
| $P_{O_2}^m$ | permeability of $O_2$ in Nafion® 112 membrane, mol $cm^{-1}\ s^{-1}$ |
| Q | charge density, C $cm^{-2}$ (geometric area) |
| $\overline{Q}$ | charge density of Pt, C $cm^{-2}$ (geometric area) |
| R | universal gas constant, 8.314 J $mol^{-1}\ K^{-1}$ |
| T | temperature, K |
| t | time, s |
| V | potential, V vs. SHE |
| v | scan rate, V s$^{-1}$ |
| W | catalyst loading, mg$_{Pt}$ cm$^{-2}$ (geometric area) |

**Greek**

| | |
|---|---|
| $\alpha$ | transfer coefficient |
| $\delta$ | Pt/C electrode thickness, cm |
| $\delta_f$ | Nafion® film thickness, cm |
| $\rho$ | density of Nafion®, g cm$^{-3}$ |
| $\nu$ | kinematic viscosity, cm$^2$ s$^{-1}$ |
| $\eta$ | overpotential, V vs. SHE |
| $\lambda$ | moles of water per sulphonic acid group in Nafion® |
| $\chi_{H_2O_2}$ | fraction of O$_2$ reducing to H$_2$O$_2$ |
| $\omega$ | electrode rotation rate, s$^{-1}$ |

**Superscript**

| | |
|---|---|
| 0 | standard state or equilibrium |
| a | anode |
| c | cathode |

**Subscript**

| | |
|---|---|
| ad | adsorbed |
| b | backward reaction, bulk |
| D | diffusion |
| disk | Pt/Nafion® coated disc electrode |
| f | Nafion® film or forward reaction |
| kin | kinetic |
| ring | Pt ring electrode |

**Acknowledgements**

The US Department of Energy supported this work under contract number DOE-DE-FC36-02AL, for which the authors are grateful. The authors thank Michael E. Fortin for oxygen permeability measurements and Shruti Modi for guidance on Fenton tests.

## Tables

**Table 1:** Average $H_2O_2$ selectivity and formation rates on supported Pt, binary and ternary Pt fuel cell catalysts. $H_2O_2$ selectivity and formation rates were respectively calculated according to equations [1] and [2]. High surface area Ketjen Black carbon (BET ~800 $m^2$ $g^{-1}$) was the support for all the catalysts unless noted otherwise.

| Catalyst | % $H_2O_2$ at 0.6 V vs. SHE | % $H_2O_2$ at 0.025 V vs. SHE | $k_f^0$, x $10^5$ $mol^2$ $cm^{-5}$ $s^{-1}$ 0-0.2 V vs. SHE | $k_f^0$, x $10^5$ $mol^2$ $cm^{-5}$ $s^{-1}$ 0.2-0.45V vs. SHE | $k_f^0$, x $10^5$ $mol^2$ $cm^{-5}$ $s^{-1}$ 0.45-0.6 V vs. SHE | $H_2O_2$ formation rates[d], mol $cm^{-2}$ $s^{-1}$ Anode[b] x $10^{12}$ | $H_2O_2$ formation rates[d], mol $cm^{-2}$ $s^{-1}$ Cathode[c] x $10^6$ |
|---|---|---|---|---|---|---|---|
| Pt | 0.75 | 2.78 | 1.87 | 2.73 | 2.56 | 0.342 | 0.007 |
| Pt[a] | 0.69 | 5.5 | 1.74 | 2.94 | 1.05 | 0.674 | 0.004 |
| PtCo | 2.11 | 5.36 | 1.34 | 14.36 | 4.17 | 0.657 | 0.017 |
| PtCo[a] | 3.65 | 8.4 | 1.84 | 1.78 | 5.21 | 1.031 | 0.022 |
| PtNi | 2.53 | 10.62 | 2.76 | 14.99 | 3.39 | 1.302 | 0.014 |
| PtRh | 0.74 | 1.28 | 2.465 | 5.15 | 1.25 | 0.157 | 0.005 |
| PtRu | 1.85 | 0.55 | 45.10 | 13.53 | 4.06 | 0.067 | 0.017 |
| PtV | 2.99 | 8.65 | 3.91 | 10.45 | 3.98 | 1.061 | 0.016 |
| PtIrCo | 5.39 | 5.05 | 6.17 | 27.18 | 9.28 | 0.338 | 0.018 |
| PtRhCo | 2.09 | 2.76 | 2.21 | 14.48 | 4.37 | 0.618 | 0.038 |

a – supported on Vulcan XC-72R

b – Evaluated at local potential at the anode was assumed to be 0 V vs. SHE, which translates to an overpotential of 0.695 V for $H_2O_2$ formation

c - local potential at the anode was assumed to be 0.6 V vs. SHE, which translates to an overpotential of 0.095 V for $H_2O_2$ formation

d – $H_2O_2$ rates were estimated at 75 °C, 1 atm and 94% RH conditions. Measured activation energies corresponding to $H_2O_2$ formation on supported Pt catalyst was used to estimate $H_2O_2$ rates at 75 °C for all supported Pt binary and ternary alloys catalysts.

**Table 2: Parameters used in the analysis of measured current at the Pt ring[a].**

| Parameters | Value | Comments |
|---|---|---|
| a | 1 | Measured, Ref. 2 |
| A | 0.164025 cm$^2$ | Ref. 16 |
| b | 2 | Measured, Ref. 2 |
| $C^*_{O_2}$ | 1.274 mol cm$^{-3}$ | Ref. 14[a] |
| $D^*_{O_2}$ | 2.2 x 10$^{-6}$ cm$^2$ s$^{-1}$ | Ref. 56 |
| $E^0$ | 0.695 V vs. SHE | - |
| EW | 1100 g mol$^{-1}$ | Ref. 57 |
| F | 96485 C mol equiv.$^{-1}$ | Ref. 58 |
| N | 0.2 | Measured |
| $\overline{Q}$ | 210 µC cm$^{-2}$ (real area) | Ref. 18, 19 |
| R | 8.314 J mol$^{-1}$ K$^{-1}$ | Ref. 58 |
| $T^0$ | 298 K | Measured |
| v | 0.01 V s$^{-1}$ | |
| α | 0.5 | Assumed |
| $\delta_f$ | 10$^{-5}$ cm | Ref. 59 |
| ρ | 1 g cm$^{-3}$ | Ref. 58 |
| υ | 0.009 cm$^2$ s$^{-1}$ | Estimated |
| ω | 2500 s$^{-1}$ | Measured |

[a] The mole fraction solubility $X_1$ of oxygen in water is given as $\ln X_1 = A_2 + \frac{B_2}{T^*} + C_2 \ln T^*$, where, $T^* = \frac{T}{100} K$. $A_2$ = -66.7354, $B_2$ = 87.4755 and $C_2$ = 24.4526.

**Table 3: (a)** Crystallite size of Pt,. PtCo and PtIrCo catalysts before and after potential cycling tests at 65 °C and 120 °C. **(b)** Relative performances of these there catalysts during cycling tests.

**(a)**

| Catalyst | Initial size (Å) | 2800 cycles 65 °C (Å) | 2800 cycles 120 °C (Å) |
|---|---|---|---|
| Pt | 33 | 45 | 61 |
| PtCo | 62 | 66 | 53 (Pt) <br> 81 (PtCo) |
| PtIrCo | 59 | 51 | 39 (Pt) <br> 60 (PtIrCo) |

**(b)**

| Catalyst | Pt | PtCo | PtIrCo |
|---|---|---|---|
| Cyclic stability | →→→→→→→→→→→→→→→→→→ | | |
| Pt concentration in membrane | ←←←←←←←←←←←←←←←←←← | | |
| Particle size | →→→ | = | = |
| Extent of sintering | ←←←←←←←←←←←←←←←←←← | | |

# Figure Captions

**Figure 1:** Schematic of a single-sided MEA with reference electrode. H2 was fed on one side while O2 was fed on the other. The O2 side had the catalyst (Pt, PtCo or PtIrCo) and was subjected to a fixed potential of 0.6 V vs. the H2 on Au reference electrode. The reference electrode was electronically isolated from the flow-field by a thin Nafion® 111 membrane.

**Figure 2:** (a) % H2O2 formed during the ORR on binary alloy catalysts PtCo (□), PtRh (○), PtRu (△), PtNi (●), PtV (■) compared to Pt (■). (b) Polarization curves for the ORR on binary alloy catalysts PtCo (dash), PtRh (dot), PtRu (dash dot), PtNi (dash dot dot), PtV (short dash) compared to Pt (line). The thin film RRDE (2500 rpm) experiments were done in 1M HClO4 solution (pH = -0.3) bubbled with pure O2 at 25 °C and 1 atm.

**Figure 3:** (a) % H2O2 formed during the ORR on ternary alloy catalysts PtIrCo (●) and PtRhCo (○) compared to Pt (■) and PtCo (□). (b) Polarization curves for the ORR on ternary alloy catalysts PtIrCo (dot), and PtRhCo (dash dot) compared to Pt/KB (line) and PtCo (dash). The thin film RRDE (2500 rpm) experiments were done in 1M HClO4 solution (pH = -0.3) bubbled with pure O2 at 25 °C and 1 atm.

**Figure 4:** Stability of binary and ternary Pt catalysts during cycling between 0.65 and 1.2 V vs. SHE in 0.1 M H2SO4 and at 25 °C. PtCo and PtIrCo, the most durable of these catalysts lost 44.58% and 44.94% of their initial electrochemical area respectively after 20000 cycles. Pt lost 85.28% of its initial electrochemical area under similar conditions.

**Figure 5:** Mass activity and durability of Pt binary and ternary catalysts supported on Ketjen black carbon relative to Pt. The durability reported here reflects the electrochemical area at the end of 20000 cycles between 0.65 and 1.2 V (5 seconds each) in 0.1 M HClO4 at 25 °C and 1 atm.

**Figure 6:** Electrochemical rate constant for H2O2 formation, $k_f$, as a function of overpotential, $\eta = E_{app} - E_0$, $E_0 = 0.695$ V vs. SHE, for Pt, PtCo and PtIrCo catalysts supported on Ketjen black carbon. The data between η values 0-0.2V, 0.2-0.45 and 0.45-0.6V was fit with three separate linear equations (shown as lines). The value of $\alpha = 0.05$, $T_0 = 25$ °C. Data was obtained in 1M HClO4 bubbled with pure O2 at 1 atm.

**Figure 7:** Estimated rates of H2O2 formation / mol cm-2 s-1 in the cathode side of a PEM fuel cell for Pt (dash dot), PtCo (dot) and PtIrCo (line) catalysts as a function of relative humidity at 75 °C. Local potential at the cathode was assumed to be ~0.6 V vs. SHE, which translates to an overpotential of 0.095 V for H2O2 formation.

**Figure 8:** Estimated rates of H2O2 formation / mol cm-2 s-1 in the anode side of a PEM fuel cell for Pt (dash dot), PtCo (dot) and PtIrCo (line) catalysts as a function

of relative humidity at 75 °C. Local potential at the cathode was assumed to be ~0 V vs. SHE, which translates to an overpotential of 0.695 V for $H_2O_2$ formation.

Figure 9: Estimated parasitic fuel loss / % as a function of $H_2O_2$ selectivity on a catalyst in the anode (line) and the cathode (dash) of a fuel cell. The fuel loss was calculated at 75 °C and 94% RH in a fuel cell operating at 2 A cm-2 with stoichiometric amounts of $H_2$ and $O_2$ while the local anode and cathode overpotentials are 0 and 0.6 V vs. SHE respectively.

Figure 10: Fluorine emission rates / μmol hr-1 measured as a function of time from single sided MEAs. The electrode with Pt (-●-), PtCO (-○-), and PtIrCo (-■-) was held at 600 mV vs. Au reference electrode throughout the test.

Figure 11: Electron microprobe analysis (EPMA) for elemental Pt, Co and Ir from the MEAs after 2800 cycles on cathode between 0.87 and 1.05 V vs. SHE (1 minute each) at 120 °C and 50% RH with $H_2$ on the anode and $N_2$ on the cathode.

Figure 12: Electrochemical area (normalized to its initial value) of the working electrode (Pt, PtCo, and PtIrCo supported on Ketjen Black) measured as a function of cycle number. The electrodes were cycled between 0.87 and 1.05 V vs. SHE (30s at each potential) at 50% RH and 120 °C. The total metal loadings were similar.

Figure 13: Weight loss (%) observed on Nafion® 111 membrane samples (2"x2") after 96 hours of Fenton testing for various concentrations of Fenton ions ($Co^{2+}$ and $Fe^{2+}$). Cobalt nitrate [$Co(NO_3)_{2.6H_2O}$] and iron sulfate ($FeSO_4.6H_2O$) were respectively used to impregnate $Co^{2+}$ and $Fe^{2+}$ into Nafion membrane.

# Figures

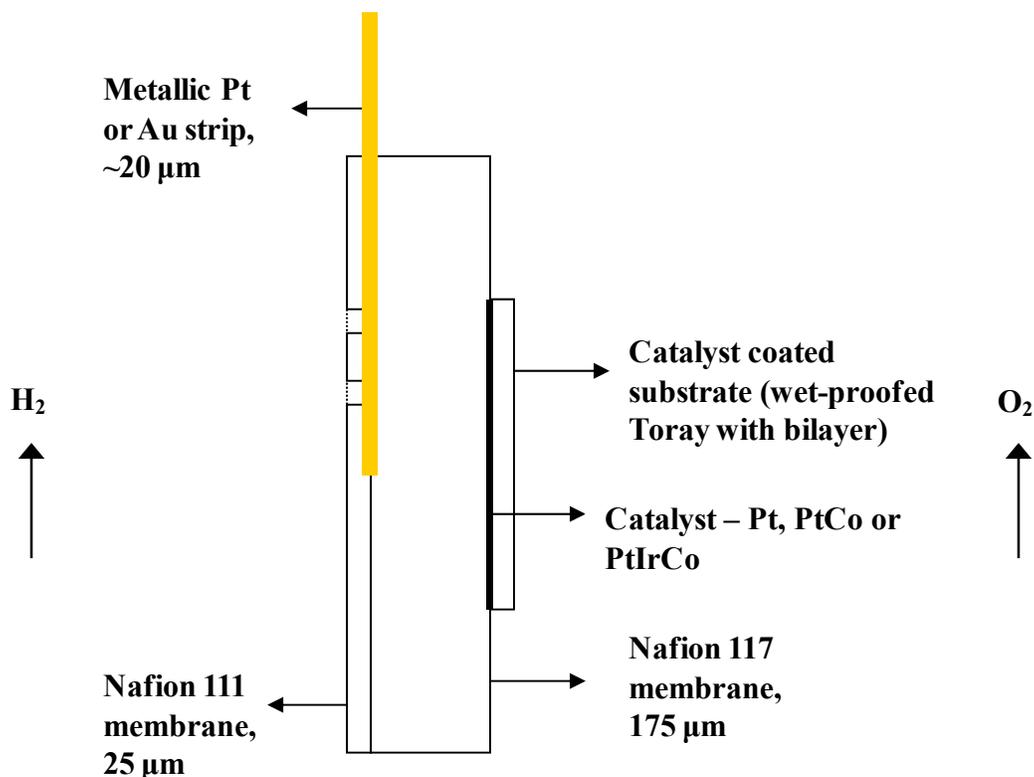

Figure 1: Schematic of a single-sided MEA with reference electrode. $H_2$ was fed on one side while $O_2$ was fed on the other. The $O_2$ side had the catalyst (Pt, PtCo or PtIrCo) and was subjected to a fixed potential of 0.6 V vs. the $H_2$ on Au reference electrode. The reference electrode was electronically isolated from the flow-field by a thin Nafion® 111 membrane.

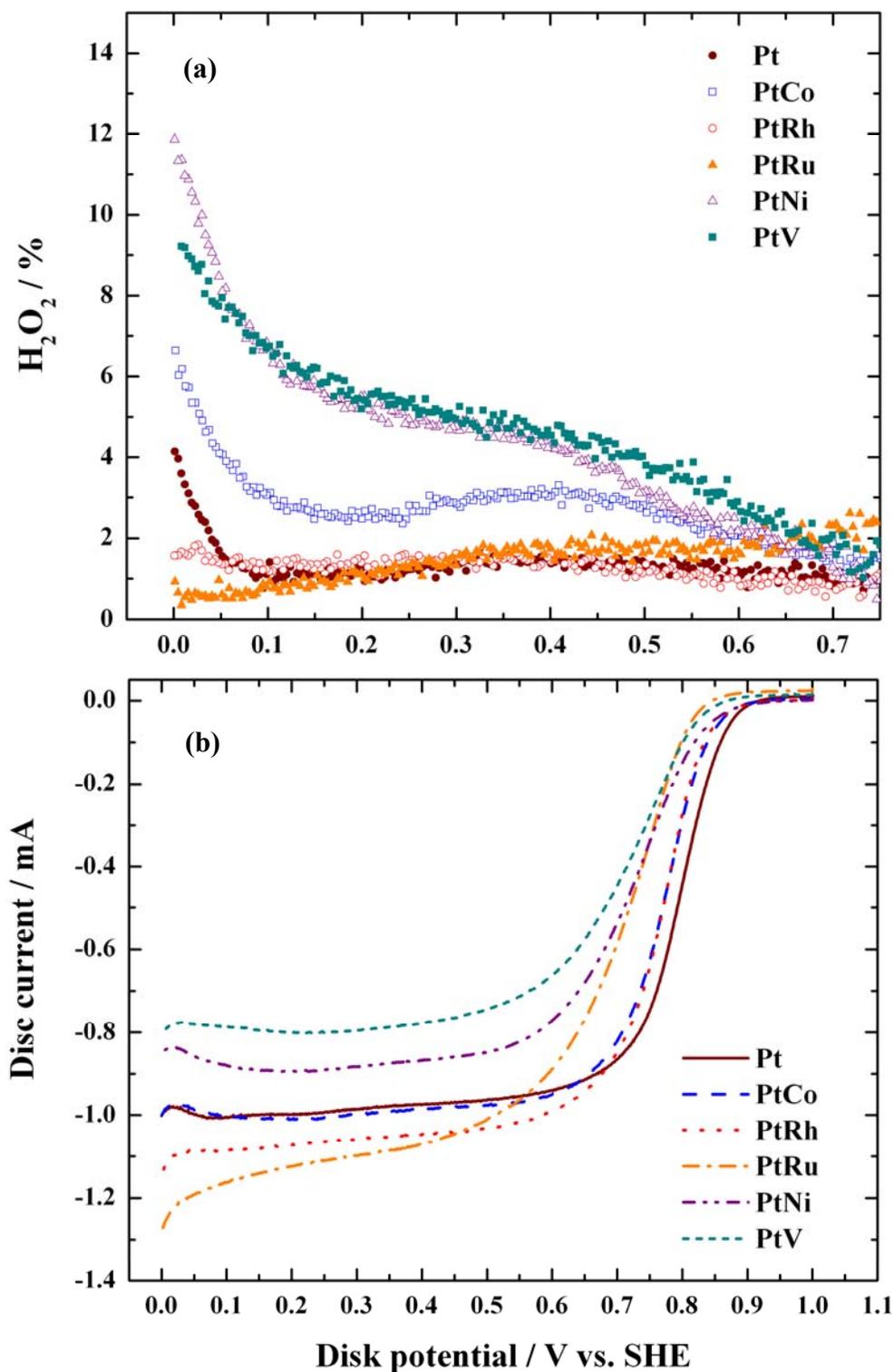

Figure 2: (a) % H$_2$O$_2$ formed during the ORR on binary alloy catalysts PtCo (□), PtRh (○), PtRu (△), PtNi (●), PtV (■) compared to Pt (■). (b) Polarization curves for the ORR on binary alloy catalysts PtCo (dash), PtRh (dot), PtRu (dash dot), PtNi (dash dot dot), PtV (short dash) compared to Pt (line). The thin film RRDE (2500 rpm) experiments were done in 1M HClO$_4$ solution (pH = -0.3) bubbled with pure O$_2$ at 25 °C and 1 atm.

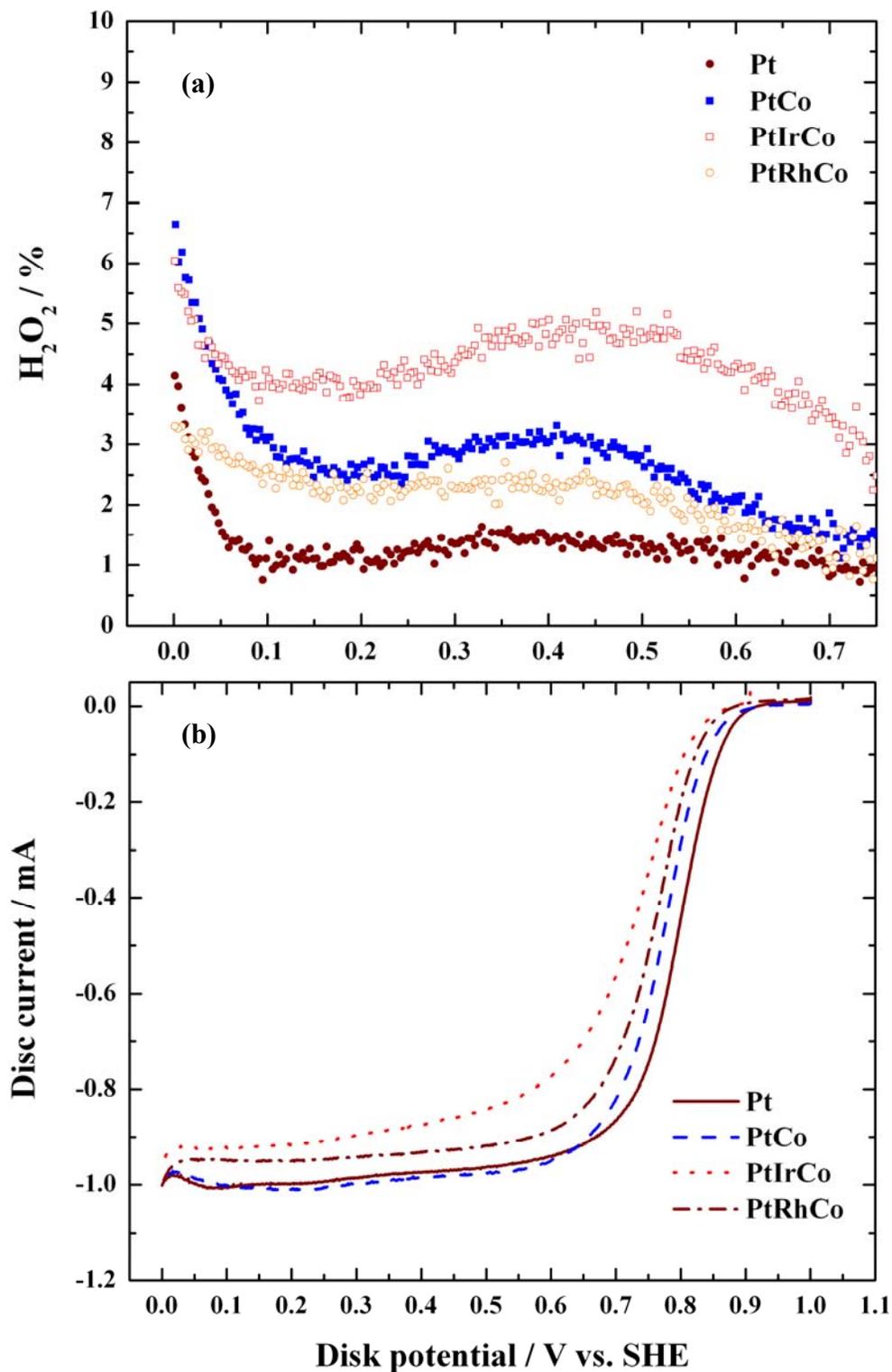

Figure 3: (a) % H$_2$O$_2$ formed during the ORR on ternary alloy catalysts PtIrCo (●) and PtRhCo (○) compared to Pt (■) and PtCo (□). (b) Polarization curves for the ORR on ternary alloy catalysts PtIrCo (dot), and PtRhCo (dash dot) compared to Pt/KB (line) and PtCo (dash). The thin film RRDE (2500 rpm) experiments were done in 1M HClO$_4$ solution (pH = -0.3) bubbled with pure O$_2$ at 25 °C and 1 atm.

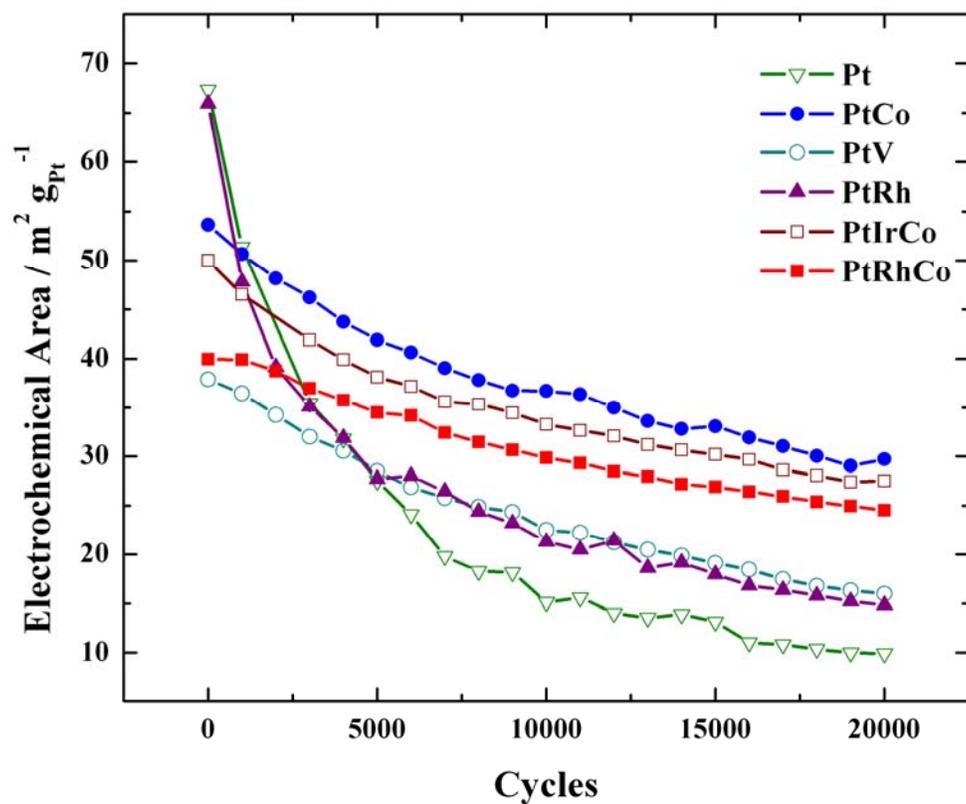

Figure 4: Stability of binary and ternary Pt catalysts during cycling between 0.65 and 1.2 V vs. SHE in 0.1 M $H_2SO_4$ and at 25 °C. PtCo and PtIrCo, the most durable of these catalysts lost 44.58% and 44.94% of their initial electrochemical area respectively after 20000 cycles. Pt lost 85.28% of its initial electrochemical area under similar conditions.

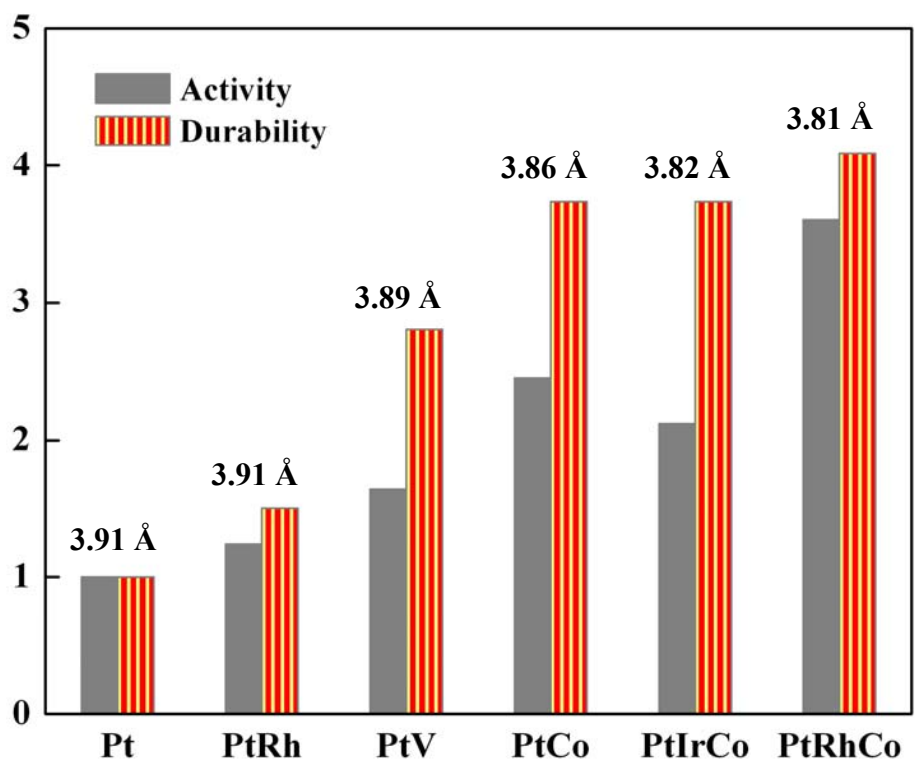

Figure 5: Mass activity and durability of Pt binary and ternary catalysts supported on Ketjen black carbon relative to Pt. The durability reported here reflects the electrochemical area at the end of 20000 cycles between 0.65 and 1.2 V (5 seconds each) in 0.1 M $HClO_4$ at 25 °C and 1 atm. The respective lattice constant value is shown above the bars for each catalyst.

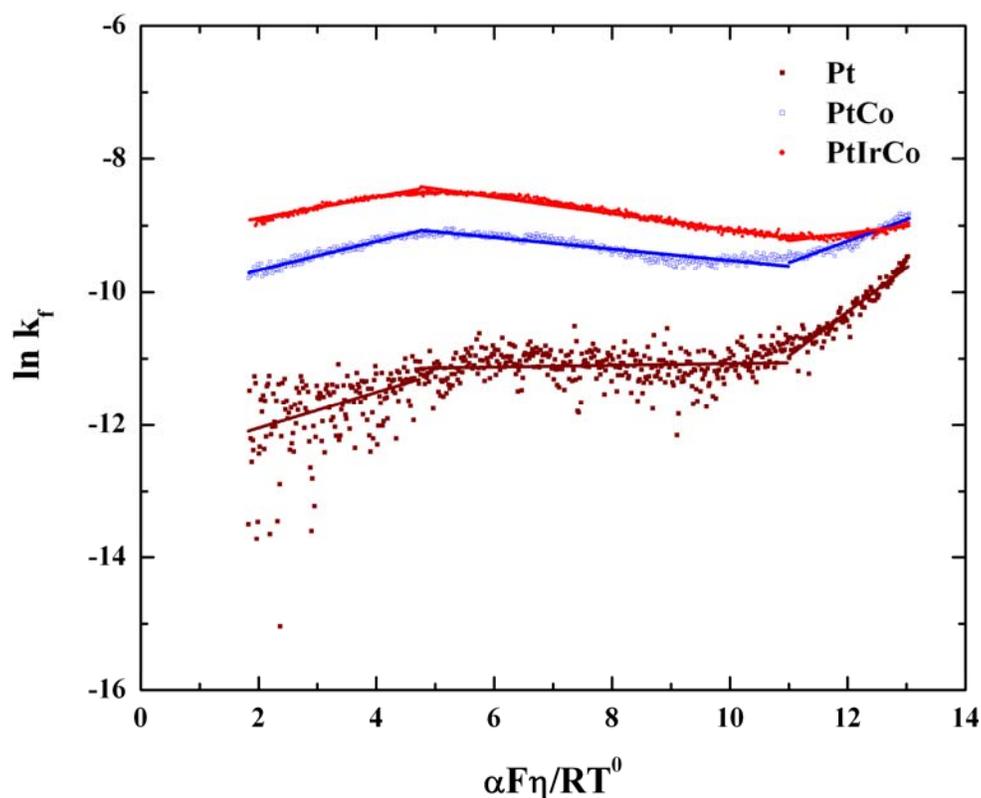

Figure 6: Electrochemical rate constant for $H_2O_2$ formation, $k_f$, as a function of overpotential, $\eta = E_{app} - E^0$, $E^0 = 0.695$ V vs. SHE, for Pt, PtCo and PtIrCo catalysts supported on Ketjen black carbon. The data between $\eta$ values 0-0.2V, 0.2-0.45 and 0.45-0.6V was fit with three separate linear equations (shown as lines). The value of $\alpha = 0.05$, $T^0 = 25$ ºC. Data was obtained in 1M $HClO_4$ bubbled with pure $O_2$ at 1 atm.

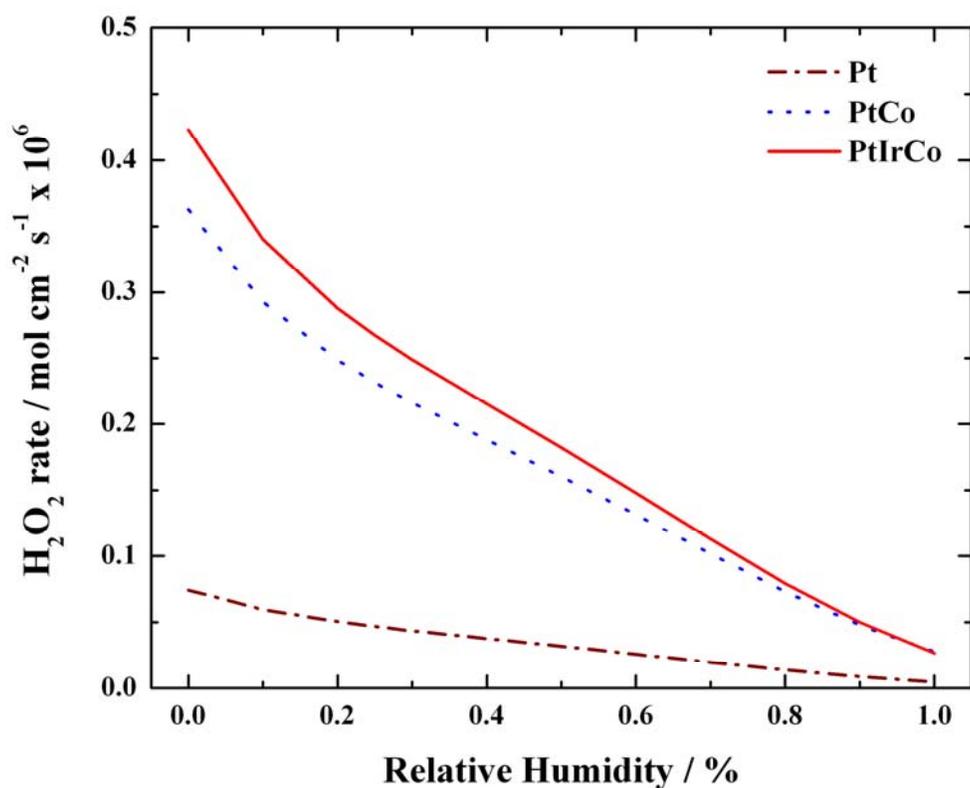

Figure 7: Estimated rates of $H_2O_2$ formation / mol cm$^{-2}$ s$^{-1}$ in the cathode side of a PEM fuel cell for Pt (dash dot), PtCo (dot) and PtIrCo (line) catalysts as a function of relative humidity at 75 °C. Local potential at the cathode was assumed to be ~0.6 V vs. SHE, which translates to an overpotential of 0.095 V for $H_2O_2$ formation.

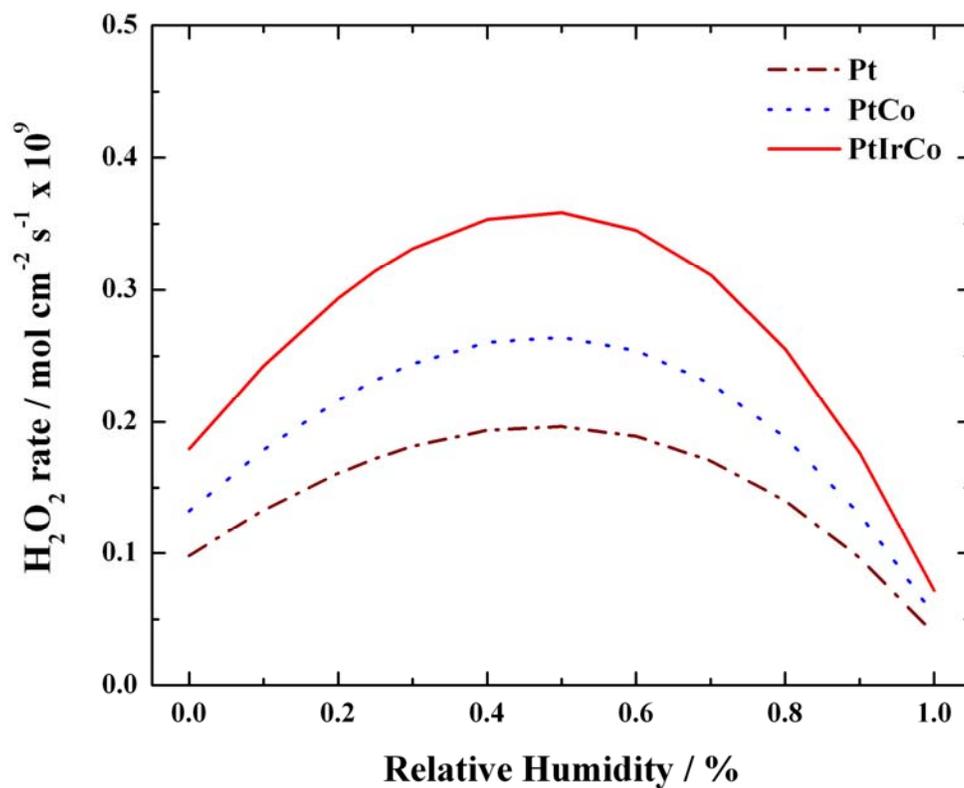

Figure 8: Estimated rates of $H_2O_2$ formation / mol cm$^{-2}$ s$^{-1}$ in the anode side of a PEM fuel cell for Pt (dash dot), PtCo (dot) and PtIrCo (line) catalysts as a function of relative humidity at 75 °C. Local potential at the cathode was assumed to be ~0 V vs. SHE, which translates to an overpotential of 0.695 V for $H_2O_2$ formation.

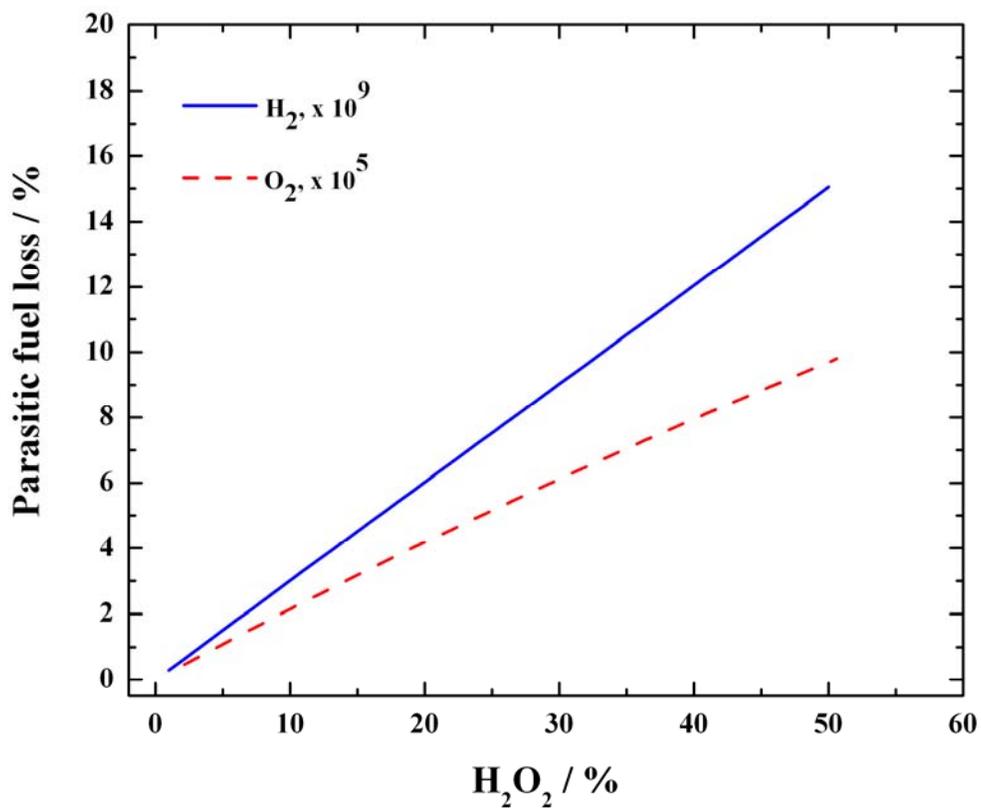

Figure 9: Estimated parasitic fuel loss / % as a function of $H_2O_2$ selectivity on a catalyst in the anode (line) and the cathode (dash) of a fuel cell. The fuel loss was calculated at 75 °C and 94% RH in a fuel cell operating at 2 A cm$^{-2}$ with stoichiometric amounts of $H_2$ and $O_2$ while the local anode and cathode overpotentials are 0 and 0.6 V vs. SHE respectively.

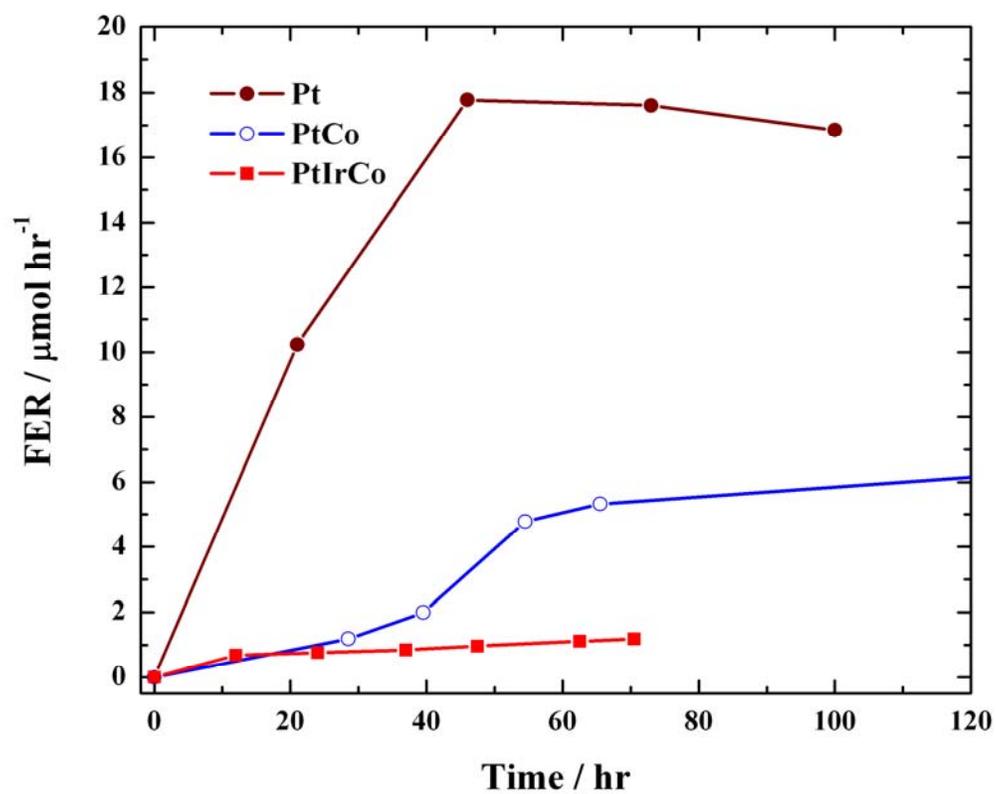

Figure 10: Fluorine emission rates / μmol hr$^{-1}$ measured as a function of time from single sided MEAs. The electrode with Pt (-●-), PtCO (-○-), and PtIrCo (-■-) was held at 600 mV vs. Au reference electrode throughout the test.

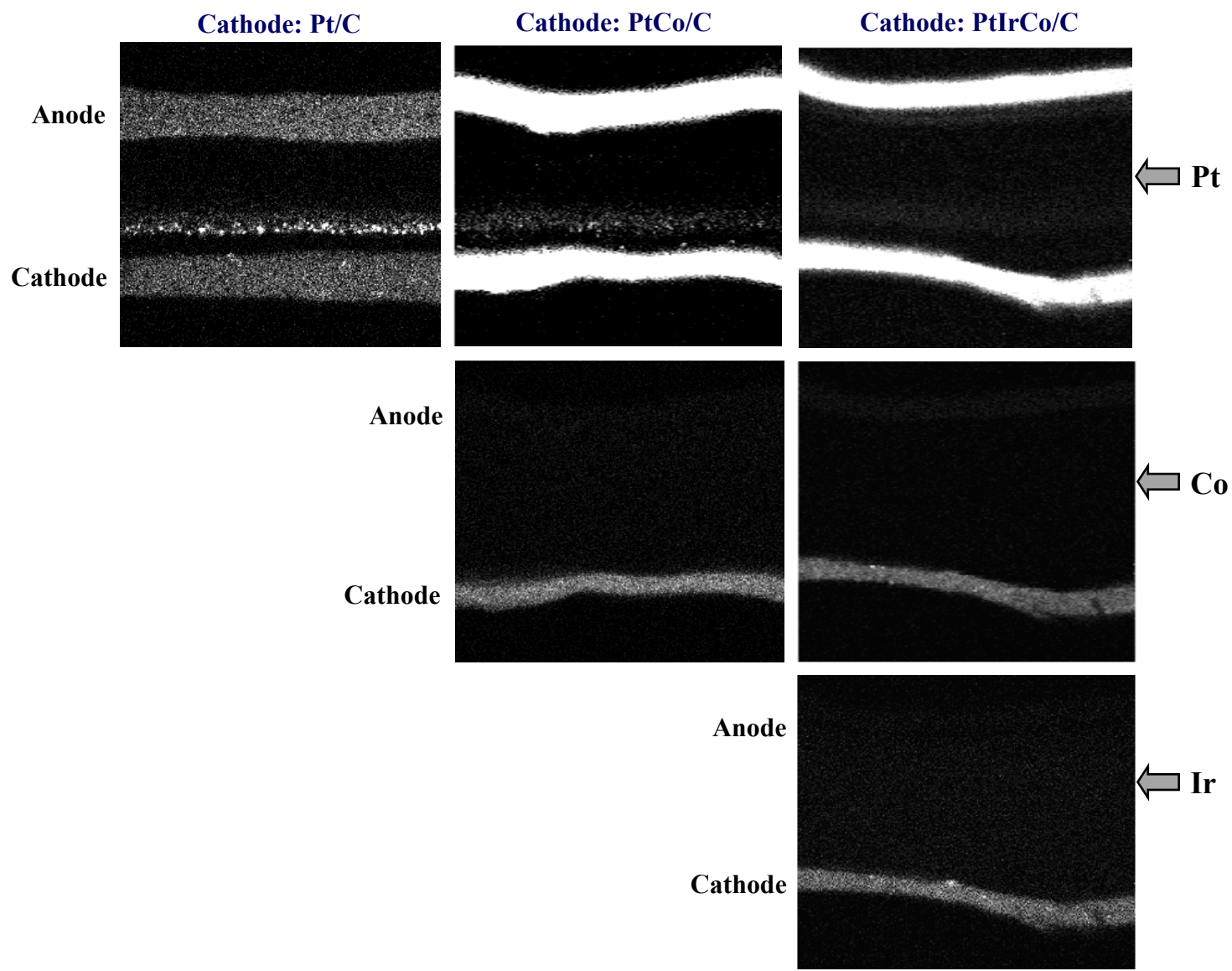

Figure 11: Electron microprobe analysis (EPMA) for elemental Pt, Co and Ir from the MEAs after 2800 cycles on cathode between 0.87 and 1.05 V vs. SHE (1 minute each) at 120 °C and 50% RH with $H_2$ on the anode and $N_2$ on the cathode.

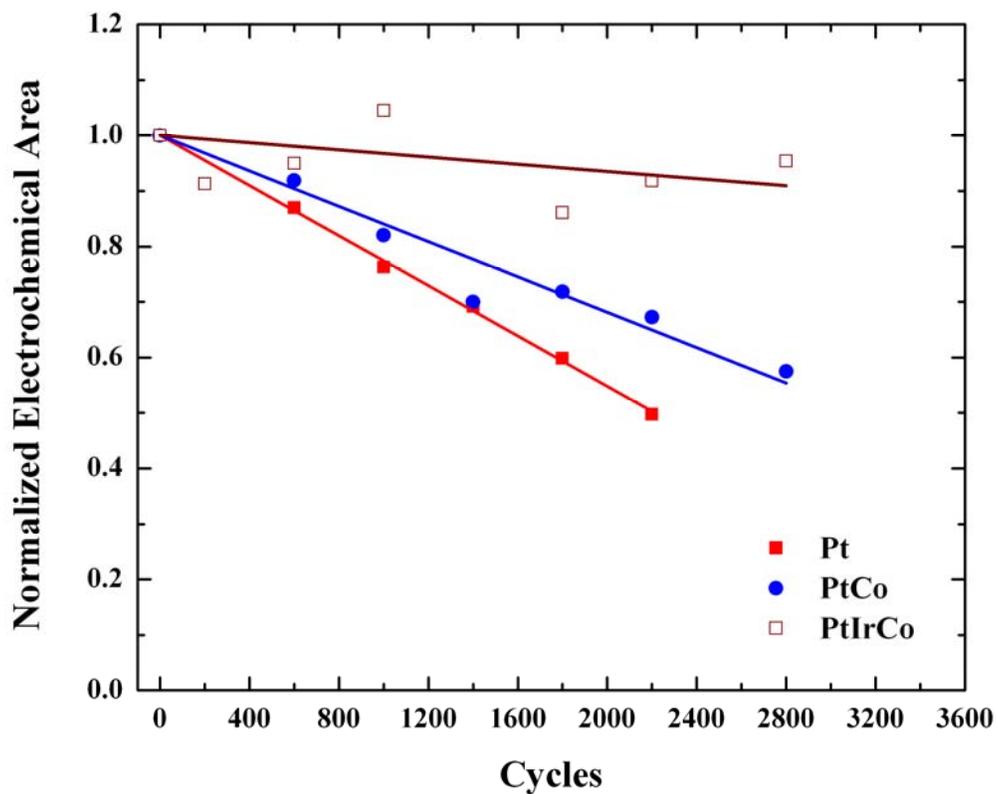

Figure 12: Electrochemical area (normalized to its initial value) of the working electrode (Pt, PtCo, and PtIrCo supported on Ketjen Black) measured as a function of cycle number. The electrodes were cycled between 0.87 and 1.05 V vs. SHE (30s at each potential) at 50% RH and 120 °C. The total metal loadings were similar.

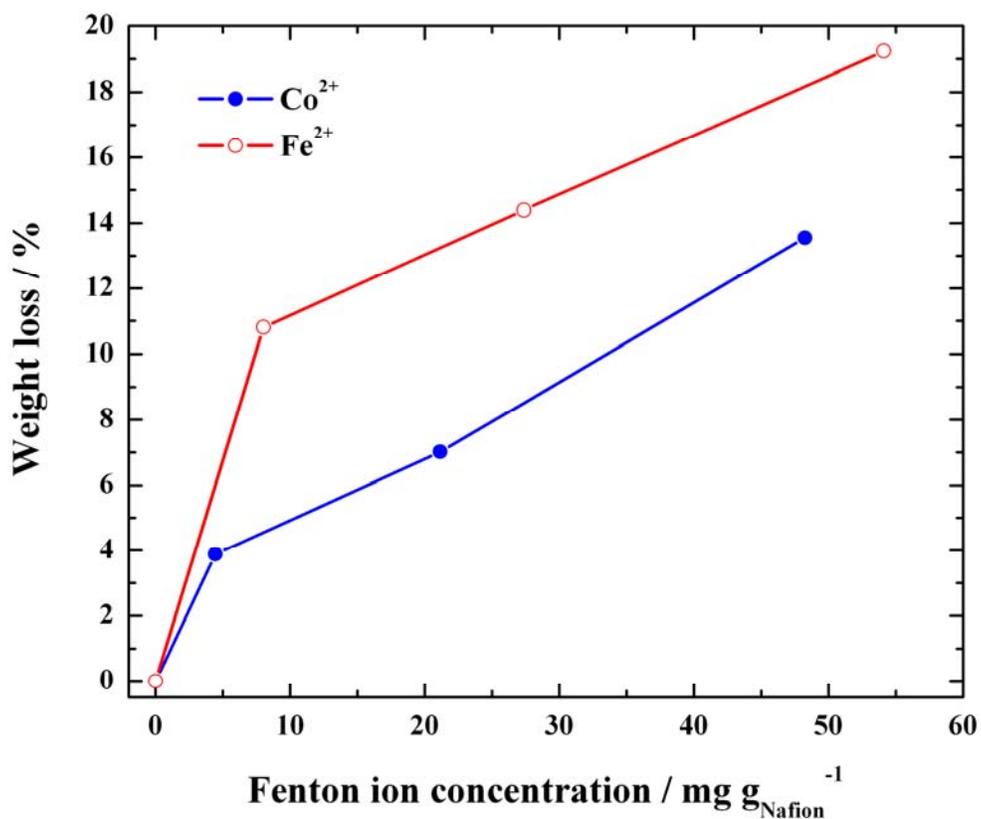

Figure 13: Weight loss (%) observed on Nafion® 111 membrane samples (2"x2") after 96 hours of Fenton testing for various concentrations of Fenton ions ($Co^{2+}$ and $Fe^{2+}$). Cobalt nitrate [$Co(NO_3)_2 \cdot 6H_2O$] and iron sulfate ($FeSO_4 \cdot 6H_2O$) were respectively used to impregnate $Co^{2+}$ and $Fe^{2+}$ into Nafion membrane.

**References**

1. E. Yeager, Electrochim. Acta 29 (1984) 1527.
2. V. A. Sethuraman, J. W. Weidner, A. T. Haug, S. Motupally, and L. V. Protsailo, J. Electrochem. Soc. 155 (2008) B50.
3. A. Bosnjakovic, and S. Schlick, J. Phys. Chem. B. 108 (2004) 4332.
4. M. Inaba, T. Kinumoto, M. Kiriake, R. Umebayashi, A. Tasaka, and Z. Ogumi, Electrochim. Acta 51 (2006) 5746.
5. M. Inaba, H. Yamada, R. Umebayashi, M. Sugishita, and A. Tasaka, Electrochemistry 75 (2007) 207.
6. T. Kinumoto, M. Inaba, Y. Nakayama, K. Ogata, R. Umebayashi, A. Tasaka, Y. Iriyama, T. Abe, and Z. Ogumi, J. Power Sources 158 (2006) 1222.
7. D. A. Schiraldi, J. Macromol. Sci. C: Polymer Rev. 46 (2006) 315.
8. US Patents 5,013,618 (1991); 4,880,711 (1989); 4,447,506 (1984).
9. US Patents 5,189,005 (1993); 5,593,934 (1997).
10. T. J. Schmidt, H. A. Gasteiger, G. D. Stäb, P. M. Urban, D. M. Kolb and R. J. Behm, J. Electrochem. Soc. 145 (1998) 2354.
11. T. J. Schmidt, U. A. Paulus, H. A. Gasteiger and R. J. Behm, J. Electroanal. Chem. 508 (2002) 41.
12. N. Markovic and P. N. Ross, J. Electroanal. Chem. 330 (1992) 499.
13. V. Stamenkovic, N. M. Markovic and P. N. Ross, Jr., J. Electroanal. Chem. 500 (2001) 44.
14. L. H. Gevantman, in: CRC Handbook of Chemistry and Physics, 79$^{th}$ ed., D. R. Lide, Editor, CRC Press, New York, 1998-99, p. 8-86.
15. W. J. Albery and M. L. Hitchman, Ring-Disc Electrodes, Clarendon Press, Oxford, 1971.
16. Rotated Ring-Disk Electrodes: DT21 Series, Pine Instrument Company, Raleigh, NC 27617.
17. S. Trasatti, and O. A. Petrii, Pure and Appl. Chem. 63 (1991) 711.
18. T. Biegler, D. A. J. Rand, and R. Woods, J. Electroanal. Chem. 29 (1971) 269.
19. J. Bett, K. Kinoshita, K. Routsis and P. Stonehart. J. Catal. 29 (1973) 160.
20. V. Mittal, H. R. Kunz, and J. M. Fenton, Abstract #1192, 208$^{th}$ Meeting of the Electrochemical Society, Los Angeles, CA, October 16, 2005.
21. V. O. Mittal, H. R. Kunz, and J. M. Fenton, Electrochem. Solid State Lett. 9 (2006) A299.
22. V. O. Mittal, H. R. Kunz, and J. M. Fenton, J. Electrochem. Soc., 153 (2006) A1755.
23. V. O. Mittal, H. R. Kunz, and J. M. Fenton, J. Electrochem. Soc., 154 (2007) B652.
24. M. R. Tarasevich, A. Sadkowski, and E. Yeager, in Comprehensive Treatise in Electrochemistry, Chapter. 6. J. O'. M. Bockris, B. E. Conway, E. Yeager, S. U. M. Khan, and R. E. White, Editors, Plenum Press, New York, 1983.
25. S. K. Zečević, J. S. Wainright, M. H. Litt, S. Lj. Gojković and R. F. Savinell, J. Electrochem. Soc. 144 (1997) 2973.
26. A. Damjanovic and G. Hudson, J. Electrochem. Soc. 135 (1988) 2269.
27. U. A. Paulus, A. Wokaun, G. G. Scherer, T. J. Schmidt, V. Stamenkovic, V. Radmilovic, N. M. Markovic and P. N. Ross, J. Phys. Chem. B. 106 (2002) 4181.

28. K. C. Neyerlin, W. Gu, J. Jorne and H. A. Gasteiger, J. Electrochem. Soc. 153 (2006) A1955.
29. A. J. Bard and L. R. Faulkner, Electrochemical Methods, Wiley & Sons, Inc., New York, 1980.
30. A. B. Anderson and T. V. Albu, J. Electrochem. Soc. 147 (2000) 4229.
31. R. A. Sidik and A. B. Anderson, J. Electroanal. Chem. 528 (2002) 69.
32. Y. Wang and P. B. Balbuena, J. Chem. Theory Comput. 1 (2005) 935.
33. V. Stamenkovic, B. N. Grgur, P. N. Ross, and N. M. Markovic, J. Electrochem. Soc. 152 (2005) A277.
34 N. A. Anastasijevic, Z. M. Dimitrijevic, and R. R. Adzic, Electrochim. Acta 31 (1986) 1125.
35. D. R. Morris and X. Sun, J. Appl. Polym. Sci. 50 (1993) 1445.
36. D. Rivin, C. E. Kendrick, P. W. Gibson, N. S. Schneider, Polymer 42 (2001) 623.
37. N. H. Jalani, P. Choi and R. Datta, J. Membrane Sci. 254 (2005) 31.
38. T. Zawodzinski, M. Neeman, L. Sillerud and S. Gottesfeld, J. Phys. Chem. 95 (1991) 6040.
39. J. T. Hinatsu, M. Mizuhata and H. Takenaka, J. Electrochem. Soc. 141 (1994) 1493.
40. L. M. Onishi, J. M. Prausnitz, and J. Newman, J. Phys. Chem. B. 111 (2007) 10166.
41. S. Motupally, A. J. Becker and J. W. Weidner, J. Electrochem. Soc. 147 (2000) 3171.
42. T. Fuller, Ph. D. Dissertation, University of California, Berkeley, 1989.
43. R. C. Reid, J. M. Prausnitz and T. K. Sherwood, The Properties of Gases and Liquids, McGraw Hill Inc., New York, 1977.
44. T. Sakai, H. Takenaka and E. Torikai, J. Electrochem. Soc. 133 (1986) 88.
45. K. Broka and P. Ekdunge, J. Appl. Electrochem. 27 (1997) 117.
46. W. Liu, and D. Zukerbrod, J. Electrochem. Soc. 152 (2005) A1165.
47. S. Burlatsky, T. Jarvi, and V. Atrazhev, Abstract # 502, The Electrochemical Society Meeting Abstracts, Volume 2005-2, Los Angeles, California, October 16-21, 2005.
48. A. Panchenko, H. Dilger, J. Kerres, M. Hein, A. Ullrich, T. Kaz and E. Roduner, Phys. Chem. Chem. Phys. 6 (2004) 2891.
49. A. Panchenko, *Dipl.-Chem.*, Institut für Physikalische Chemie der Universität Stuttgart, October 2004.
50. P. Yu, M. Pemberton, and P. Plasse, J. Power Sources 144 (2005) 11.
51. L. Protsailo, Development of High Temperature Membranes and Improved Cathode Catalysts, United States Department of Energy Report, December 2005.
52. A. T. Haug, L. V. Protsailo, S. Modi, V. A. Sethuraman, and S. Motupally, "Characterization of Membranes and Catalysts for High Temperature PEM Fuel Cells", United States Department of Energy - High Temperature Working Group, 208[th] Meeting of the Electrochemical Society, Los Angeles, CA, October 20, 2005.
53. V. A. Sethuraman, J. W. Weidner, A. T. Haug, and L. V. Protsailo, J. Electrochem. Soc. 155 (2008) B119.
54. Z. Jusys, and R. J. Behm, J. Phys. Chem. B 108 (2004) 7893.
55. M. Inaba, M. Sugishita, J. Wada, K. Matsuzawa, H. Yamada, and A. Tasaka, J. Power Sources 178 (2008) 699.
56. Q. Guo, V. A. Sethuraman, and R. E. White, J. Electrochem. Soc. 151 (2004) A983.
57. K. A. Mauritz and R. B. Moore, Chemical Reviews, 104 (2004) 4535.

58. R. H. Perry and D. W. Green, Perry's Chemical Engineer's Handbook, 7$^{th}$ Edition, McGraw Hill, 1997.
59. U. A. Paulus, T. J. Schmidt, H. A. Gasteiger and R. J. Behm, J. Electroanal. Chem. 495 (2001) 134.